\documentclass[journal]{IEEEtran}
\usepackage{amsmath,amsfonts}
\usepackage{algorithmic}
\usepackage{algorithm}
\usepackage{array}
\usepackage[caption=false,font=normalsize,labelfont=sf,textfont=sf]{subfig}
\usepackage{textcomp}
\usepackage{stfloats}
\usepackage{url}
\usepackage{verbatim}
\usepackage{graphicx}
\usepackage{cite}
\usepackage{booktabs}
\usepackage{tabularx}
\usepackage{multirow}
\usepackage{tikz}
\usepackage{pifont}
\usepackage{orcidlink}

\begin{document}
\bstctlcite{IEEEexample:BSTcontrol}

\title{EdgeALM: An Efficient and Unified Edge Accelerator for Large Audio Language Models via Full-Stack Hardware-Software Co-Design and Agent-Driven Recursive Optimization}
\author{Chenglin Yang$^{\orcidlink{0009-0007-7421-6938}}$, Feifan Xu, Xingyuan Li, Xinpeng Xing, and Dihu Chen$^{\orcidlink{0000-0001-5432-8149}}$
\thanks{This work was funded by the Key Area R \& D Program of Guangdong Province under Grant 2022B0701180001.
\emph{(Corresponding authors: Xinpeng Xing; Dihu Chen.)}

Chenglin Yang, Feifan Xu, Xingyuan Li, Xinpeng Xing, and Dihu Chen are with the School of Integrated Circuits,
Sun Yat-sen University, Shenzhen 518107, China
(e-mail: yangchlin5@mail2.sysu.edu.cn, xuff8@mail2.sysu.edu.cn, lixy688@mail2.sysu.edu.cn, xingxp@mail.sysu.edu.cn, stscdh@mail.sysu.edu.cn).
}}
\maketitle

\begin{abstract}
  Edge deployment of audio language models (ALMs) is constrained by
  large model footprints, heterogeneous workloads, and limited
  software support. This paper presents EdgeALM, an efficient and
  unified FPGA accelerator for ALMs that combines full-stack
  hardware-software co-design with agent-driven recursive optimization.
  An end-to-end ALM compression flow is developed, combining a unified
  AWQ-style quantizer with modality-conditioned vocabulary pruning.
  A Conv-Matmul Disaggregation architecture is proposed, with
  dedicated matrix and convolution engines that operate concurrently
  on their respective workloads.
  An Extra-Coarse-Grained ISA is designed to incorporate regular
  on-chip data movement and execution sequences into compute
  instructions, reducing software complexity with modest hardware
  control overhead.
  Hardware and software recursive loops are introduced for
  agent-driven full-stack optimization. 
  The proposed EdgeALM is implemented on a single Alveo U50 in an
  ordinary desktop PC and demonstrates inference on more than 20
  models, including ALMs with up to 13B parameters within the 8-GB
  HBM capacity.
  It achieves high-throughput audio-to-text inference at up to
  169.5 tokens/s and sustains real-time streaming audio-to-audio
  inference,
  with 94\% bandwidth utilization and 26~W power consumption on average.
  Standalone ASR, text-only LLM, and TTS models are also well
  supported, and can be combined in cascades. 
  Compared with state-of-the-art FPGA accelerators (e.g., EdgeLLM), 
  EdgeALM scales to larger models and supports a wider range of inference workloads extended to the audio modality, 
  while achieving higher bandwidth utilization at lower board power and cost.
\end{abstract}


\begin{figure}[!t]
\centering
\includegraphics[width=\columnwidth]{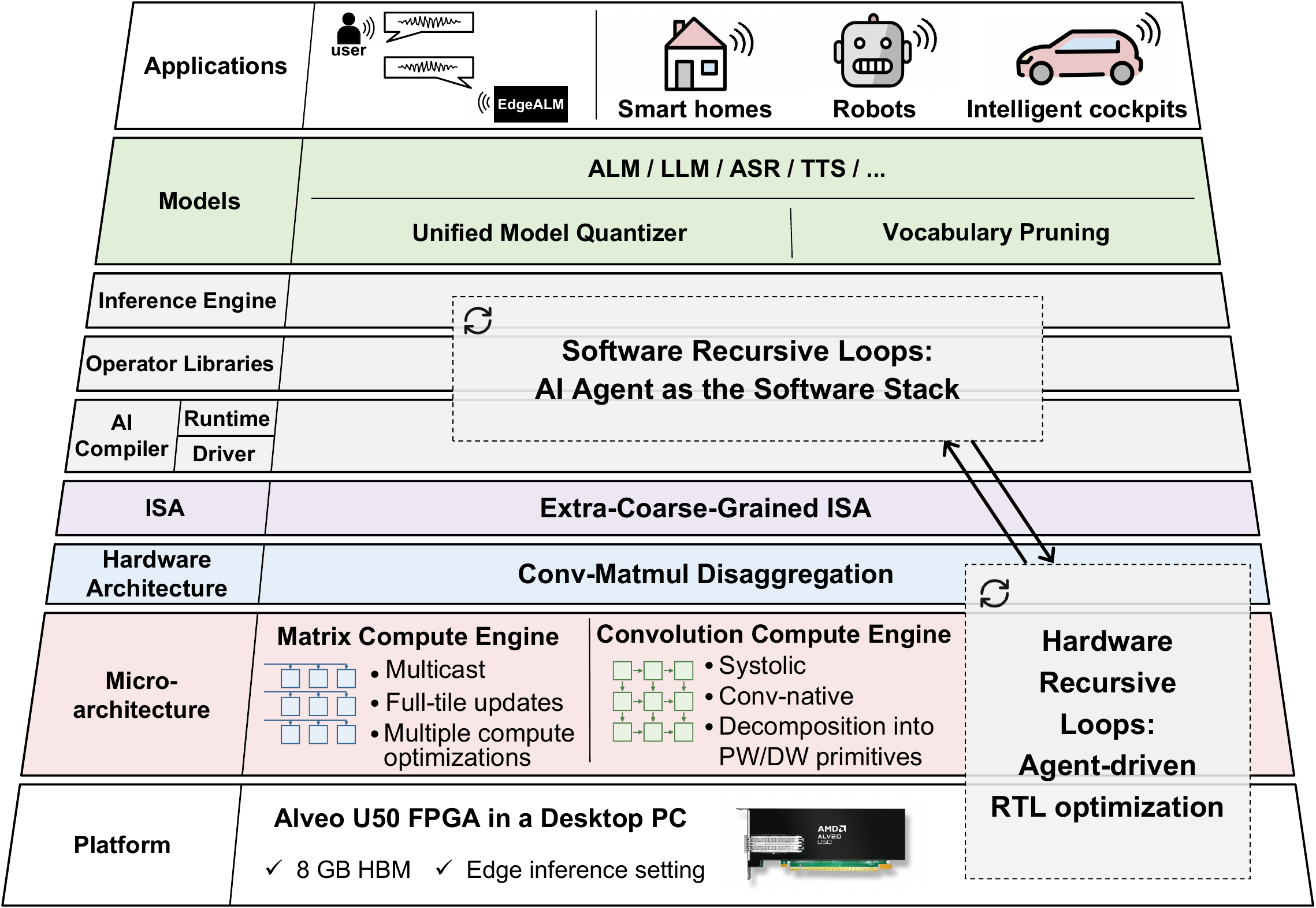}
\caption{Overview of the full stack for LLM/ALM inference and the proposed EdgeALM's full-stack design.}
\label{fig:overview}
\end{figure}
\begin{figure*}[!t]
\centering
\includegraphics[width=\textwidth]{"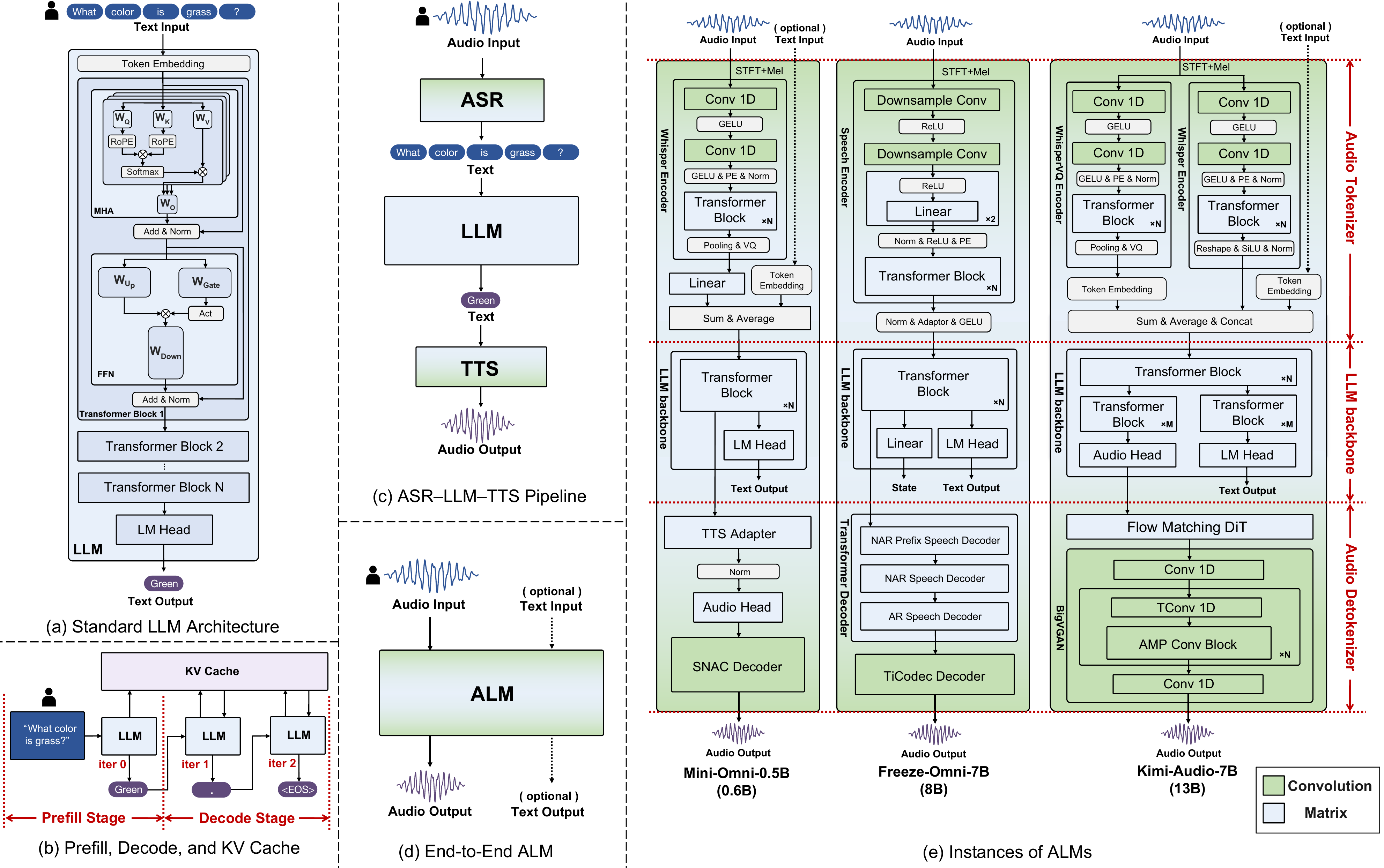"}
\caption{From Text-only LLMs to End-to-End Audio Language Models.}
\label{fig:LLM}
\end{figure*}

\section{Introduction}

Large language models (LLMs) have achieved remarkable progress and are now moving toward edge devices, 
where interaction with users and the physical world spans audio, vision, and other modalities. 
These models have accordingly evolved beyond text-only interaction into multimodal variants such as vision-language models (VLMs), 
vision-language-action models (VLAs), and audio language models (ALMs). 
Of all modalities, audio is the most natural, direct, and efficient 
medium for human-machine interaction, and has therefore found broad application in intelligent cockpits, smart homes, and robotics.
Conventional voice interaction systems typically adopt an ``ASR-LLM-TTS'' cascade: 
automatic speech recognition (ASR) transcribes speech into text, an LLM understands the text and generates a text response, 
and text-to-speech (TTS) converts the text back into speech. With text as the only interface between modules, 
recognition errors propagate and accumulate along the pipeline. Paralinguistic information such as emotion 
and tone is also lost, degrading the naturalness of interaction.
ALMs, or Large Audio Language Models (LALMs), also known as Speech Language Models (SLMs) or Spoken Language Models (SLMs), 
break this text bottleneck by unifying the pipeline end to end. 
A typical ALM also consists of three modules: an audio tokenizer, an LLM backbone, and an audio detokenizer. 
These modules exchange continuous or discrete speech tokens directly, without any text-only intermediate. 
This design therefore preserves speech information and improves naturalness.

However, ALM deployment on edge devices remains challenging due to steep hardware resource demands and an immature software ecosystem. 
Take Kimi-Audio~\cite{kimiaudio}, a 13B-parameter ALM with a 7B LLM backbone, as an example: On the hardware side, its released weights have a total size of 40 GB, 
and
require 48 GB of GPU memory. NPUs and non-CUDA platforms are explicitly excluded from the official support roadmap. 
On the software side, 
the official inference scripts do not support streaming inference. 
Even vLLM~\cite{vllm}, a widely used inference engine, took nearly a year from the community's initial 
request to its official support for Kimi-Audio, and only for speech understanding, not speech generation. 
It is also nontrivial to fully quantize them to low-bit precision for edge deployment because such ALMs integrate multiple heterogeneous modules.
These obstacles span the hardware and software stacks, calling for full-stack optimization.

For the hardware deployment of ALMs, GPUs are the dominant platform. 
However, their high power consumption and cost exceed the budgets of edge devices. 
Edge voice interaction is also inherently single-user and streaming, and since autoregressive decoding with a batch size of one is memory-bound, 
the massive compute capability of GPUs remains largely idle and energy efficiency low. 

FPGAs offer a promising alternative, providing high memory bandwidth and a better compute-to-bandwidth ratio 
for the memory-bound workload at the edge. They also allow the reconfigurable fabric to be tailored 
to the characteristics of specific workloads, and the large on-chip memory to be fully utilized to effectively reduce off-chip accesses. 
Several FPGA-based LLM accelerators have been proposed recently~\cite{flightllm,edge_llm,clat,accllm,pushlimit,hummingbird,bape}.
However, prior FPGA-based LLM accelerators have demonstrated limited model scale and generality, 
and have not yet addressed the demands of edge ALM inference:
1) They are designed for text-only LLMs and do not match the workloads of ALMs;
2) the models they support are limited to 8B parameters at most in the dense case;
3) they each demonstrate inference with only one or two LLMs, providing no evidence of generality across diverse models; and
4) their platforms generally sit at two extremes, as cards such as the Alveo U280 are designed for data center acceleration 
and must be installed in rack or tower servers, making them less well suited to edge deployment, while embedded boards 
such as the Kria KV260 lack both the bandwidth for real-time speech interaction and the compute to keep latency low.
In addition, existing AI accelerators for audio tasks~\cite{liou_attention_asr,yang_rsnn_asr,tambe_sm6,liu_twn_asr,u_sparta}
focus only on small, task-specific speech models and do not support ALMs.
To address these challenges, this paper presents the following key contributions:
\begin{itemize}
\item An end-to-end ALM compression flow is developed, combining
a unified AWQ-style quantizer with modality-conditioned vocabulary
pruning. 

\item A Conv-Matmul (CM) Disaggregation architecture is proposed, with specialized
matrix and convolution engines that can operate concurrently.
The matrix engine matches its throughput to the HBM bandwidth
through full-array weight updates and efficient DSP use, while
the convolution engine natively supports diverse 1D convolution
operators without Img2col.

\item An Extra-Coarse-Grained ISA is designed to incorporate
regular on-chip data movement and execution sequences into
compute instructions, reducing software complexity with modest
hardware control overhead. A hierarchical descriptor mechanism
further reuses instruction sequences across tiles and reduces
host-launch overhead.

\item Software and hardware recursive loops are introduced for
agent-driven full-stack optimization. The two software loops
construct and optimize deployment software from scratch, without
established compiler infrastructure, inference engines, or a
predefined IR. The two hardware loops refine an existing RTL design
for timing closure and inference performance. Both sides exchange
information to guide further optimization.

\item  Implemented on a single Alveo U50 in an ordinary desktop PC, 
EdgeALM supports a range of ALMs with up to 13B
parameters, as well as ASR, 
LLM, and TTS models. It delivers
high-throughput audio-to-text and real-time streaming audio-to-audio
inference, with bandwidth utilization of 94\% at a
power of 26~W on average.
\end{itemize}

\section{Background and Related Work}
\subsection{LLMs and ALMs}
An LLM generates text token by token, and is typically built
as a decoder-only Transformer~\cite{vaswani}. It maps input text into hidden
states with a token embedding table, processes them with a stack of
Transformer blocks, and projects the result onto the vocabulary with an LM
head, producing the distribution of the next token. As depicted in
Fig.~\ref{fig:LLM}(a), each Transformer block
consists of multi-head attention (MHA) and a feed-forward network
(FFN), and both are composed mainly of linear projections,
so the computation of an LLM is dominated by dense matrix multiplications.
Inference with an LLM proceeds in two stages, prefill and decode, as
illustrated in Fig.~\ref{fig:LLM}(b). In the prefill stage, the model
processes the entire prompt in parallel, and the key and value tensors
computed at every layer are kept as the KV cache for the decode steps that
follow. In the decode stage, the model generates tokens one at a time.
Each generation step reads all model weights, but uses each weight only once.
Decoding is therefore a
memory-bound workload. This property is further amplified at the batch size
of one in single-user edge deployment, and it largely shapes the accelerator
design presented in this paper.

ALMs extend LLMs with the ability to listen and speak.
Voice interaction is conventionally realized by an
``ASR-LLM-TTS'' cascade (Fig.~\ref{fig:LLM}(c)), whereas ALMs integrate
understanding and generation end to end within a single model
(Fig.~\ref{fig:LLM}(d)).
An ALM consists of an audio
tokenizer, an LLM backbone, and an audio detokenizer.
The audio tokenizer converts the input waveform into discrete audio tokens or
continuous features, typically with a convolutional front end that
downsamples spectral features. The LLM
backbone is itself a decoder-only Transformer, and extends the standard LM
head with an audio head, so that the model produces both text tokens and
audio tokens. The
audio detokenizer, which also involves convolutional computation, converts the generated audio tokens back into waveforms,
either through a codec decoder, or through a
flow-matching Diffusion Transformer (DiT) followed by a
vocoder.
Fig.~\ref{fig:LLM}(e) presents three representative instances of ALMs, 
which differ in scale and module design but all
follow the organization described above. As color-coded in the figure,
convolutional computation concentrates in the audio tokenizer and the audio
detokenizer, while the LLM backbone
is dominated by matrix computation. 

\subsection{AI Accelerators for LLMs and Audio Tasks}
Several FPGA-based accelerators have been proposed for LLM inference.
FlightLLM~\cite{flightllm} is the first to demonstrate 7B LLM inference
on a single FPGA, using a configurable sparse DSP chain and an
always-on-chip decoding scheme. Its HBM bandwidth
utilization is 65.9\%, limited by pipeline bubbles.
AccLLM~\cite{accllm} further supports long-context LLM inference
through W2A8KV4 quantization, N:M semi-structured pruning, and
$\Lambda$-shaped attention. Both of them target the Alveo U280, a data-center
accelerator card intended for rack or tower servers rather than ordinary
desktop PCs.
EdgeLLM~\cite{edge_llm} employs a mixed-precision compute engine built 
around a grouped vector systolic array of customized PEs, supporting FP16$\times$FP16 
computation for MHA and FP16$\times$INT4 computation for FFN. It also achieves
an average HBM bandwidth utilization of about 75\%. 
However, a drawback of EdgeLLM is that its mixed-precision compute engine 
utilizes only 75\% of its DSP capacity in MHA mode and delivers only 
one-quarter of the peak compute available in FFN mode.
Several studies explore LLM acceleration on embedded FPGAs~\cite{pushlimit, hummingbird, hummingbirdplus}.
These embedded designs, however, sit at the opposite extreme, since their
boards provide only 19.2 to 34~GB/s of DDR4 bandwidth and very limited DSP
resources.
These platform limits would constrain extensions to real-time ALM inference, 
where audio processing must run alongside LLM generation under tight compute, 
bandwidth, and latency requirements.
All of these accelerators are built for text-only LLMs, with each evaluated on 
only one or two models and no supported dense model exceeding 8B parameters, 
as noted in Section~I. Their generality across diverse models and their 
applicability to ALM inference have not yet been demonstrated.

For audio tasks, several AI accelerators~\cite{liou_attention_asr,yang_rsnn_asr,tambe_sm6,liu_twn_asr}
have been developed, most of which support one or two of the following
tasks: ASR, speaker verification, keyword spotting, voice activity detection, speech enhancement,
speech separation, and audio classification. U-Sparta~\cite{u_sparta}
goes further by using operator decomposition to reuse a configurable
convolution engine across all these tasks.
However, these accelerators deploy only million-scale models, trading accuracy
for hardware efficiency, and support neither recent billion-scale task-specific
speech models (e.g., Qwen3-ASR~\cite{qwen3_asr}) nor speech generation tasks
such as TTS, let alone ALMs for spoken dialogue.

\subsection{The Full Software-Hardware Stack for LLM/ALM}
Figure~\ref{fig:overview} summarizes the layers involved in developing and
deploying an LLM/ALM inference edge accelerator, from the application at the top to the
hardware platform at the bottom. The upper layers \emph{Application} and \emph{Model} 
define the workload, 
and there is typically a model compressor to quantize the model 
to low-bit formats.
Moving down the stack,
the \emph{Inference Engine} (e.g., vLLM~\cite{vllm}, SGLang~\cite{sglang}, 
and llama.cpp~\cite{llamacpp}) turns the model into an executable request
workflow, coordinating prefill and decode, KV cache management, sampling,
and output streaming.
The \emph{Operator Library} then supplies reusable implementations of
operations, and the \emph{AI Compiler} maps these operations and the model
graph to a target device. For a custom accelerator, MLIR~\cite{mlir} or
TVM~\cite{tvm} can provide infrastructure for layout transformation,
tiling, fusion, memory planning, and scheduling, 
and a target-specific compiler backend, potentially leveraging LLVM~\cite{llvm},
needs to be developed to perform hardware-specific lowering and
generate code for the accelerator's Instruction Set Architecture (ISA).
Two additional concepts are the
\emph{Runtime} and \emph{Driver}, which
together support the execution of inference workloads.
Collectively, these components form a typical software stack 
for an LLM/ALM inference edge accelerator.

The \emph{ISA} serves as the interface between software and
hardware. It is an abstract model that defines the commands software
can issue
and the behavior hardware
must provide. 
The granularity of ISAs has a significant impact on both hardware
and software.
Most AI accelerators today already expose coarse-grained ISAs.
Such instructions
exploit a central regularity of AI workloads: repeated
multiply-accumulate operations can be driven by
coarse-grained control, reducing
software complexity 
without substantial control-logic overhead.
Yet they capture only part of that regularity: on-chip buffer
accesses and execution sequences may also follow certain regular
patterns in tiled computations, but this regularity is often not
exploited to the same extent, leaving their
orchestration to software and
increasing software complexity.
Tile-level abstractions then shift much of this complexity from
programmers to compilers. Triton~\cite{triton} and
TileLang~\cite{tilelang} let programmers express tiled computations
while compilers handle much of 
the lower-level mapping;
A virtual tile ISA called PTO is introduced over Ascend's existing
hardware and software interfaces.
The question, then, is whether simple hardware control could also
exploit these regular patterns, 
directly reducing software complexity
without substantially increasing hardware complexity, while reducing
repeated instruction fetch/decode/issue or instruction storage. 
This motivates our Extra-Coarse-Grained ISA, presented in
Section~\ref{sec:isa}.

Below the ISA lies the hardware stack. The \emph{Hardware Architecture}
defines the accelerator's high-level organization and dataflow, including
how its compute engines, memories, interconnects, and control interact.
The \emph{Microarchitecture} implements this organization through
concrete compute arrays, datapaths, pipelines, buffers, and controllers.
The \emph{Platform} is the physical deployment system, such as a
GPU, an FPGA, or an ASIC chip, and sets the
available resources, memory bandwidth, I/O, clock rates, and power limits.

\subsection{Agent-Driven Optimization}
Recently, AI agents have begun to contribute across the full stack.
On the hardware side, MGEMMV~\cite{mgemmv} uses multimodal LLMs to generate
Verilog from GEMM circuit diagrams, while EEsizer~\cite{eesizer} uses an LLM
agent for  closed-loop circuit optimization.
On the software side,
earlier work~\cite{llm_aided,kernelcraft,autocomp}
used LLMs and agents to automate kernel translation, generation,
optimization, and mapping. 
AI as a Compiler~\cite{ai_compiler} goes further: for selected Triton
kernels, the agent acts as the compiler, generating PTX directly
instead of using Triton's conventional optimizing and lowering path.
Sol-H3~\cite{sol_h3} explores a recursive 
self-improvement (RSI) loop to iteratively optimize
inference kernels and memory layouts.
Redwood~\cite{redwood} is an early attempt to extend agent-driven
development across hardware and software. From a human-authored
specification, its AI system generates RTL, verification collateral,
drivers, and kernels, and deploys inference
on an FPGA. Nevertheless, its evaluation remains
confined to text-only 0.6B LLM inference, leaving the
model-scale, model-diversity, and practical edge-deployment challenges
unaddressed.

\section{Model Compression for Audio Language Models}
\label{sec:quantization}
This section presents our end-to-end and hardware-aware model compression approach for ALMs.
Widely used quantization 
tools
typically support only the LLM backbone of ALMs and 
cannot quantize all modules end to end. We develop a general compression flow for ALMs: an end-to-end 
AWQ-style quantizer for heterogeneous ALM modules, together with modality-conditioned vocabulary 
pruning of the text and audio heads, 
so that the entire ALM model fits within capacity-limited DRAM (e.g., 8 GB HBM on an Alveo U50).

\subsection{Unified quantizer for heterogeneous ALM modules}
Weight-only quantization is particularly
attractive for memory-constrained inference: model weights are far larger
than activations and are repeatedly read during
autoregressive inference. Among such methods, AWQ (Activation-aware Weight
Quantization)~\cite{awq} can reduce weights to INT4 with little accuracy
loss by preserving the salient weight channels identified from activation
statistics. However, the original AWQ method
targets text-only LLMs and the W4A(FP)16 format
(INT4-weight, FP16-activation). It does not directly support ALM modules, 
nor does it support the ALM
calibration traces and the W4A(BF)16 format (INT4-weight, BF16-activation)
we ultimately adopt.
We build a unified, end-to-end AWQ-style model quantizer for ALMs that
extends the AWQ quantization interface, calibration methodology,
activation format, and scale-fusion rules to cover these heterogeneous ALM
modules under one hardware-compatible contract.
The remainder of this subsection details these techniques.

\subsubsection{Unified W4A(BF)16 Format}
\label{subsec:quant-contract}

Our flow adopts a unified format contract: weights are quantized to asymmetric INT4 with 
a group size of 128, while activations and scales remain in BF16. Temporary accumulations 
are kept in a custom 20-bit E8M11 floating-point format (1 sign bit, 8 exponent bits, 
11 mantissa bits) to preserve accuracy. This contract substantially reduces weight 
storage and HBM traffic while retaining the exponent range required by audio activations. 
Moreover, fixing a single format contract also simplifies the hardware.
The activation format is chosen as BF16 rather than FP16 based on two key observations. 
First, we find experimentally that FP16 activations overflow in audio DiT computations, 
whereas BF16 retains the FP32 exponent range and is free of this failure. Second, 
for the other two ALM modules, the audio tokenizer and the LLM, BF16 is also the 
higher-quality format on average. The audio-to-text chat task runs through exactly these two modules. 
In the evaluation of this task, BF16 achieves a higher average score than FP16, 
as summarized in Table~\ref{tab:bf16-fp16}.

\begin{table}[t]
\centering
\caption{Audio-to-text chat evaluation of Kimi-Audio on the
OpenAudioBench dataset with FP16 and BF16
activations, 
scored by a DeepSeek-V3.2-Exp judge.}
\label{tab:bf16-fp16}
\begin{tabular}{lrrr}
\toprule
\textbf{Dataset} & \textbf{Original} & \textbf{W4A(FP)16} & \textbf{W4A(BF)16} \\
\midrule
Alpaca Eval & 69.90 & 66.18 & 66.68 \\
Llama Questions & 81.00 & 81.33 & 80.00 \\
Reasoning QA & 69.21 & 66.60 & 68.22 \\
Trivia QA & 62.20 & 60.60 & 61.50 \\
Web Questions & 71.40 & 68.40 & 69.70 \\
\midrule
\textbf{Average} & 70.74 & 68.62 & 69.22 \\
\bottomrule
\end{tabular}
\end{table}

\subsubsection{Extending Calibration to Heterogeneous ALM Modules}

A key difficulty in extending AWQ to ALM modules is calibration data. Text
LLMs have broad, ready-made text calibration datasets, but ALMs cannot reuse
them directly: the input seen by an ALM's LLM backbone is created
only after an audio waveform has passed through an audio tokenizer. 
We therefore generate the
calibration set by running real audio through the deployed front end and
audio tokenizer, constructing the actual audio/text prompt, and recording the
hidden states at each quantization interface. 
For the audio detokenizer, we
similarly capture the states produced by the real trajectory rather
than using unrelated text samples.  

\subsubsection{AdaLN-aware AWQ scale fusion for audio DiT}
DiTs in ALMs are the most difficult part of these extensions: 
unlike other Transformer-based models, 
a DiT contains an additional AdaLN module. AWQ's scale-fusion rule, 
which was derived for plain Transformer blocks, does not cover this DiT-specific dataflow. 
Existing quantization methods for DiTs, such as Q-DiT~\cite{qdit} and PTQ4DiT~\cite{ptq4dit}, 
do not apply either. This is because they were designed for 
image and video diffusion models, and assume the spatial structure of image/video 
latents rather than audio semantic-token conditioning, streaming execution, 
or variable sequence lengths. We solve the quantization of the audio 
DiT by applying AWQ with both the calibration trace and the scale-fusion rule redesigned, 
as detailed below.

\begin{figure}[t]
\centering
\includegraphics[width=\columnwidth]{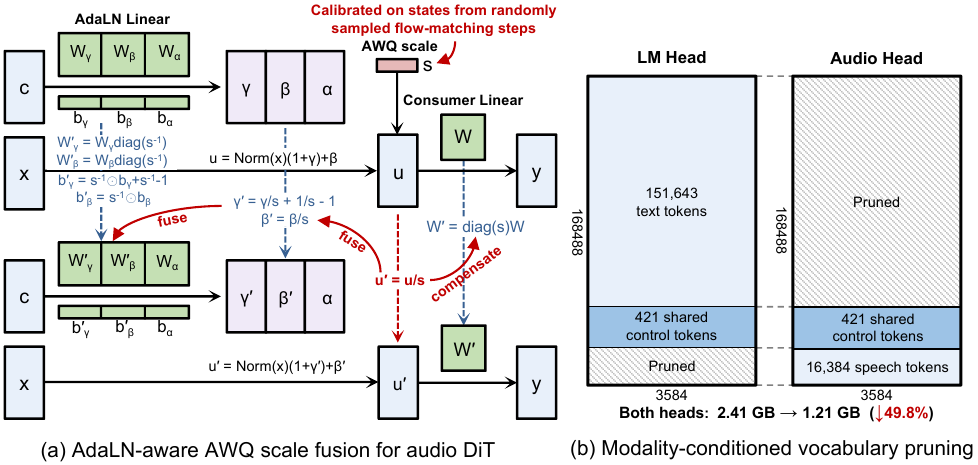}
\caption{(a) AdaLN-aware AWQ scale fusion for audio DiT.
(b) Modality-conditioned vocabulary pruning of the text and audio output heads.}
\label{fig:adaln-fusion}
\end{figure}

The ordinary AWQ fusion rule for a LayerNorm--Linear chain is not valid
after an AdaLN. An audio-DiT block generates its modulation parameters,
typically ordered as
$[\gamma,\beta,\alpha]$,
where $\gamma$ is the modulation, $\beta$ is the AdaLN shift,
and $\alpha$ is the residual gate, all produced by the quantized AdaLN linear
from the conditioning vector $c$. 
The AdaLN modulation computes
$u=\operatorname{Norm}(x)(1+\gamma)+\beta$.
To deliver an activation divided by the AWQ scale $s$ to the
consumer, the AdaLN output $u$ must be divided by $s$.
If we write
$u'=u/s=\operatorname{Norm}(x)(1+\gamma')+\beta'$, with primes denoting the
fused parameters, 
then preserving the original function requires
\begin{equation}
  \beta'=\frac{\beta}{s},\qquad
  \gamma'=\frac{\gamma}{s}+\left(\frac{1}{s}-1\right).
  \label{eq:adaln-fuse}
\end{equation}
Since $\beta$ and $\gamma$ are not stored but computed as
$\beta=cW_\beta+b_\beta$ and $\gamma=cW_\gamma+b_\gamma$, where
$W_\beta,W_\gamma$ and $b_\beta,b_\gamma$ are the corresponding weight
columns and bias slices of the AdaLN linear, the above transform is realized
by rescaling:
\begin{equation}
\begin{aligned}
  W_\beta'  &=W_\beta\,\operatorname{diag}(s^{-1}),&\quad
  b_\beta'  &=s^{-1}\!\odot b_\beta,\\
  W_\gamma' &=W_\gamma\,\operatorname{diag}(s^{-1}),&\quad
  b_\gamma' &=s^{-1}\!\odot b_\gamma+s^{-1}-\mathbf{1},
\end{aligned}
\label{eq:adaln-linear-fold}
\end{equation}
where $\odot$ denotes elementwise multiplication and $\mathbf{1}$ is the
all-ones vector.
The consumer matrix absorbs the reciprocal scale as
$W'=\operatorname{diag}(s)\,W$,
so that $u'W'=(u/s)\,\operatorname{diag}(s)\,W=uW$ and the consumer output
is preserved.
Fig.~\ref{fig:adaln-fusion} visualizes our quantization solution for audio DiT,
with $s$ obtained by calibrating on DiT states captured at 
randomly selected steps of real flow-matching trajectories.
As verified experimentally on Kimi-Audio, our quantization preserves
the perceptual quality of generated speech: the mean opinion score
(MOS) changes from 4.417 to 4.415, a decrease of only 0.002,
and the mean squared error (MSE) between the Mel spectrograms
of speech generated with and without quantization is only
$3.859\times10^{-3}$.


\subsection{Modality-conditioned vocabulary pruning}
\label{subsec:output-space-pruning}

Output heads are usually left unquantized to preserve accuracy, so they
occupy a considerable fraction of the model footprint. This effect is more
pronounced in multimodal models, which carry several modality-specific output
heads where a text-only LLM has only one: on Kimi-Audio, for example, the
BF16 text and audio heads of the LLM backbone together occupy 2.41 GB.
Since quantization usually should not be applied here, we resort to another compression
technique, namely pruning. We observe that a considerable portion of these
output heads is redundant: each head is allocated over the full joint
vocabulary, even though only the active branch's tokens, together with the 
control tokens it requires, are ever supposed to be sampled.
We therefore prune each output projection to the legal rows of its branch:
rows whose token IDs can never be generated are deleted. 
As shown in Fig.~\ref{fig:adaln-fusion}(b), this method reduces
the combined storage of the two heads from 2.41~GB to 1.21~GB,
a reduction of 49.8\%.
Because the deleted
rows are unreachable under the decoding contract, the pruning is exact: it
removes storage without approximating the distribution over legal outputs, 
and it leaves the input embedding table untouched.

\section{System Architecture}
\subsection{Overall architecture}

\begin{figure}[t]
\centering
\includegraphics[width=\columnwidth]{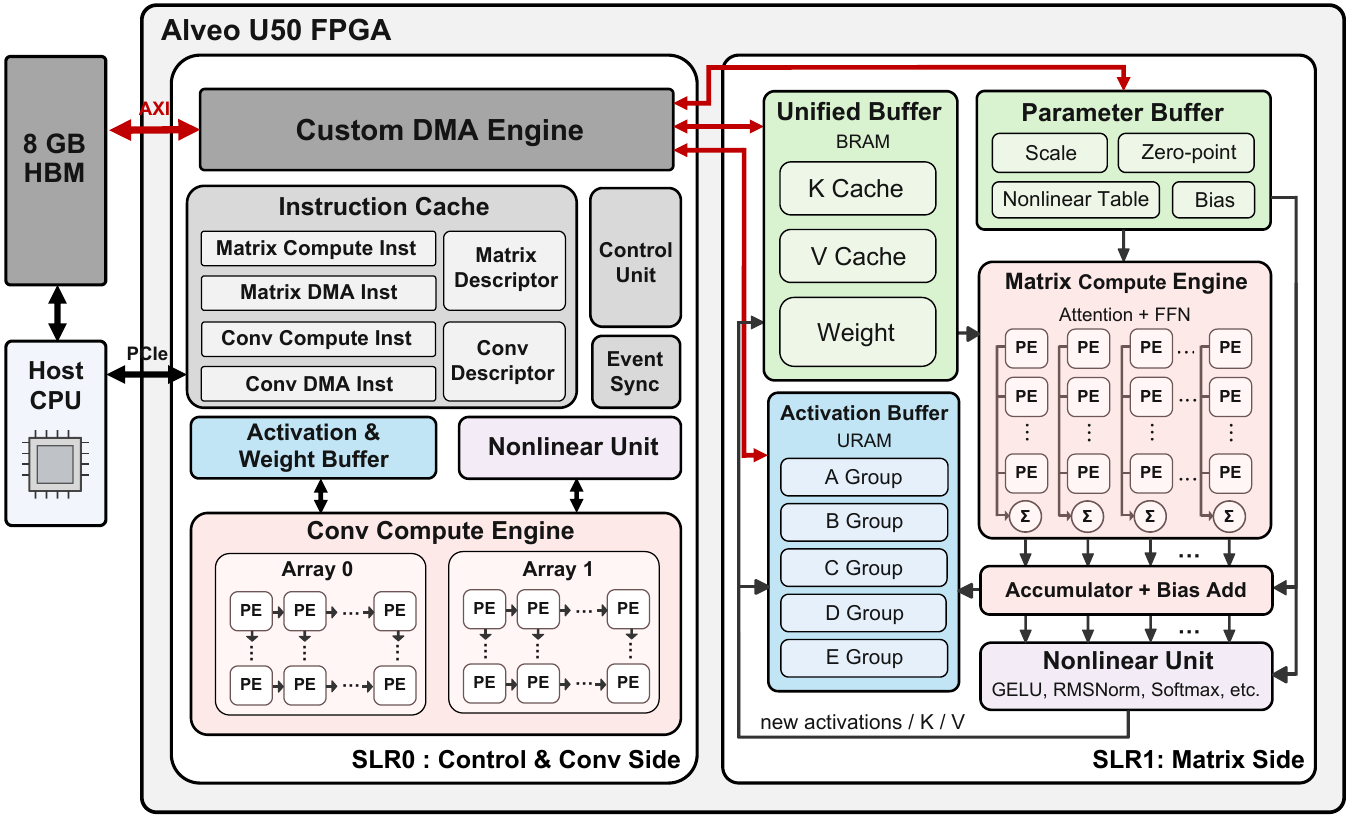}
\caption{Overall architecture of the proposed EdgeALM.}
\label{fig:edgealm-architecture}
\end{figure}

The overall architecture of EdgeALM is shown in Fig.~\ref{fig:edgealm-architecture}.
It is deployed on the Alveo U50 FPGA platform, which contains two super logic regions (SLRs) and 8~GB of HBM.
Our floorplan places the matrix compute part on SLR1,
with the control and convolution compute parts on SLR0,
forming the proposed Conv-Matmul Disaggregation architecture.

The Matrix Compute Engine (MCE) executes the matrix-dominated workloads of ALMs.
The Convolution Compute Engine (CCE) executes the convolutions.
A nonlinear unit on each side covers normalization, activation, and element-wise operations.
Multiple on-chip buffers hold the weights, KV cache, activations, and parameters consumed by the two engines.
All data movement between HBM and these buffers goes through a custom DMA engine,
whose bandwidth is shared by the two engines.
The remaining modules form the control path, which implements the proposed ISA:
the instruction cache holds the compute and DMA instructions of the two engines, together with the descriptors;
the control unit sequences these instruction streams,
and the event synchronization unit maintains the completion events
through which the streams express their data dependencies.

\subsection{U50 Platform and HBM Memory Subsystem}
\label{sec:hbm}
\begin{figure}[t]
\centering
\includegraphics[width=\columnwidth]{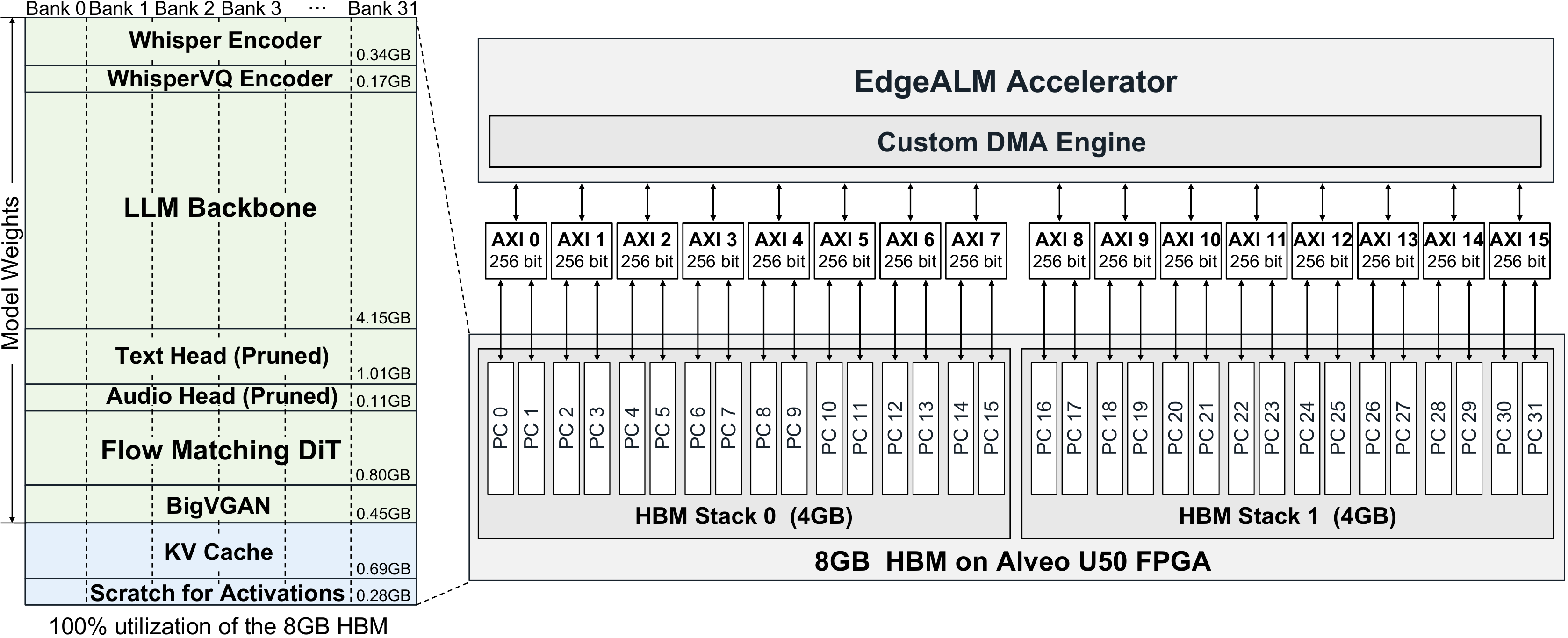}
\caption{HBM memory subsystem. Left: vertical placement of
model weights, KV cache, and activation scratch across the 32
banks, illustrated with Kimi-Audio 13B. Right: mapping of adjacent
PC pairs to the 16 256-bit AXI masters of the Custom DMA
Engine.}
\label{fig:hbm}
\end{figure}

The Alveo U50 is marketed as a data-center accelerator card.
Unlike larger data-center cards such as the Alveo U280 and
Alveo V80, which must be installed in rack or tower servers
with strong forced airflow, the U50 has a board-power limit of
only 75~W, so it can be plugged into an ordinary
desktop computer, as we do in this work. This setup brings
HBM-equipped acceleration to a genuine edge inference setting.
The U50 integrates 8 GB of HBM2 memory in two 4 GB stacks,
organized as 32 pseudo-channels (PCs), with a nominal bandwidth
of 201~GB/s.
Each PC provides an
independent 256-MB address region, called a bank. In practice,
however, the HBM bandwidth accessible to large U50 designs
falls far below the nominal value, for two reasons:
1) Placement and routing congestion.
Choi~\cite{choi2020hbm} reports that routing
congestion near the HBM ports limited their U50 design to only
24 PCs. Corda~\cite{corda_u50} likewise leaves
unused PCs between employed ones due to similar
congestion. In practice, large U50 designs~\cite{qin2024hbm, pensa}
therefore connect only a subset of the 32 PCs,
which provide only about 50~GB/s bandwidth.
However, this also utilizes only part of the 8-GB HBM
capacity, whereas LLM inference needs the full capacity
to hold its large weights.
2) HBM power limit.
The HBM2 power
budget of the U50 is constrained to
7.8~W--10~W.
Bancroft~\cite{bancroft}, although
engaging up to 30 of the 32 PCs, reports that power throttling
of the U50 HBM caps the accessible bandwidth at about 34\% of
the nominal value, with measured throughput stalling at
approximately 70~GB/s.

Under these congestion and power constraints, we design an HBM
memory subsystem that exploits the attainable bandwidth while
utilizing the entire 8-GB HBM capacity. As illustrated in
Fig.~\ref{fig:hbm}, the weights of each module are striped
across all 32 banks, with the modules stacked vertically.
The Custom DMA Engine transfers data between
the HBM and the on-chip buffers through 16 256-bit AXI master
interfaces clocked at 160~MHz, where AXI master $i$ is mapped to the adjacent
pseudo-channel pair $\mathrm{PC}_{2i}$ and $\mathrm{PC}_{2i+1}$
for $i=0,\ldots,15$, together spanning all 32 pseudo-channels
across the two 4-GB HBM stacks. In our experiments, this
pairing routed successfully with significantly reduced
congestion, whereas a one-to-one 32-master mapping failed to
route. The pairing keeps the entire 8-GB capacity accessible at
the cost of halving the peak transfer bandwidth. Nevertheless, the
resulting ceiling
$16 \times 256~\mathrm{bit} \times
160~\mathrm{MHz} = 81.92~\mathrm{GB/s}$ 
already exceeds the reported 70~GB/s
attainable by large U50 designs.

\subsection{Conv-Matmul Disaggregation}
Recent LLM inference optimization has produced disaggregated architectures
such as Prefill-Decode (PD) Disaggregation~\cite{splitwise} and
Attention-FFN (AF) Disaggregation~\cite{step3}.  Their common philosophy is
to analyze how different computational stages differ in their workload
characteristics, such as arithmetic intensity, memory access, and latency
requirements, and to place each stage on the hardware that suits it best,
typically across chips or nodes in datacenter-scale serving.

We propose Conv-Matmul (CM) Disaggregation for edge ALM inference, which
disaggregates within a single chip: a matrix engine executes the
matrix-multiplication workloads of the ALM's Transformer modules,
while a convolution-native engine executes the convolutions in
front-end speech processing and back-end speech generation.  The two engines maintain
independent compute control and local dataflows while sharing one physical
HBM DMA.
We argue that CM Disaggregation offers the following two advantages,
together with the design considerations behind them.

\subsubsection{Natural Division of Convolution and Matrix Computation in Speech Models}

In an ALM,
front-end speech
feature extraction and back-end speech generation typically rely on
convolution, whereas the LLM backbone in the middle and the optional
diffusion model are typical Transformers dominated by matrix computation.
A complete speech interaction system therefore naturally contains both
convolution and matrix computation.
The positions of the two types of computation
are concentrated
and fixed in the pipeline.  
Mapping them onto separate compute paths therefore allows each
engine to be optimized for its own characteristics: a convolution-native,
systolic-style engine that turns the high data reuse of the 1-D
convolutions in speech models into high utilization, and a non-systolic,
direct-wiring multicast engine specialized for the memory-bound matrix
computation of LLM decoding.

\subsubsection{Parallelism for Full-Duplex and Half-Duplex Interaction}
\begin{figure}[t]
\centering
\includegraphics[width=\columnwidth]{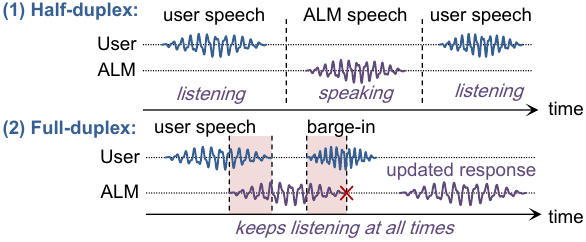}
\caption{Half-duplex versus full-duplex ALM interaction.}
\label{fig:duplex}
\end{figure}
ALMs are typically half-duplex: listening and speaking take turns, as in
Kimi-Audio~\cite{kimiaudio} and
Mini-Omni~\cite{mini_omni}.  CM Disaggregation is a natural fit
for this scenario and hides latency well: while the matrix engine
decodes the next chunk, the convolution-native engine simultaneously
generates speech for the previous chunk.  
Looking ahead, a recent line of ALMs, such as Freeze-Omni~\cite{freeze_omni}, enables
full-duplex interaction: the model keeps listening while it speaks
(Fig.~\ref{fig:duplex}).  
CM Disaggregation provides additional
opportunities to overlap matrix and convolution computation
across speech input and output, beyond pipelining consecutive
output chunks.



\begin{figure}[t]
\centering
\includegraphics[width=\columnwidth]{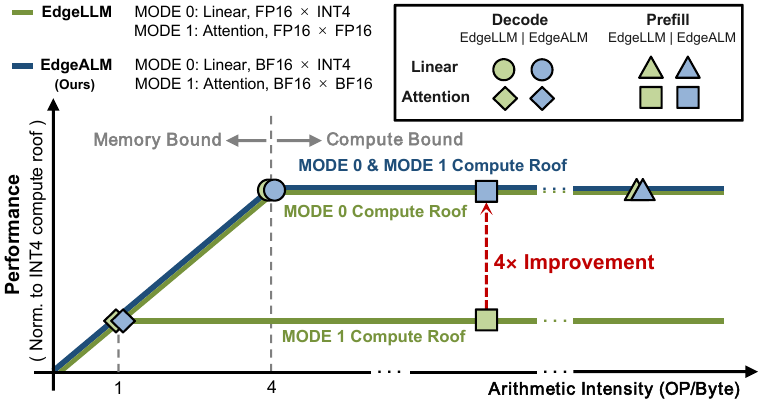}
\caption{Roofline comparison of EdgeALM and EdgeLLM~\cite{edge_llm} for linear and attention modes. Each design is normalized to its linear-mode compute roof.}
\label{fig:roofline}
\end{figure}

\subsection{Matrix Compute Engine}
\begin{figure*}[!tp]
\centering
\includegraphics[width=\textwidth]{"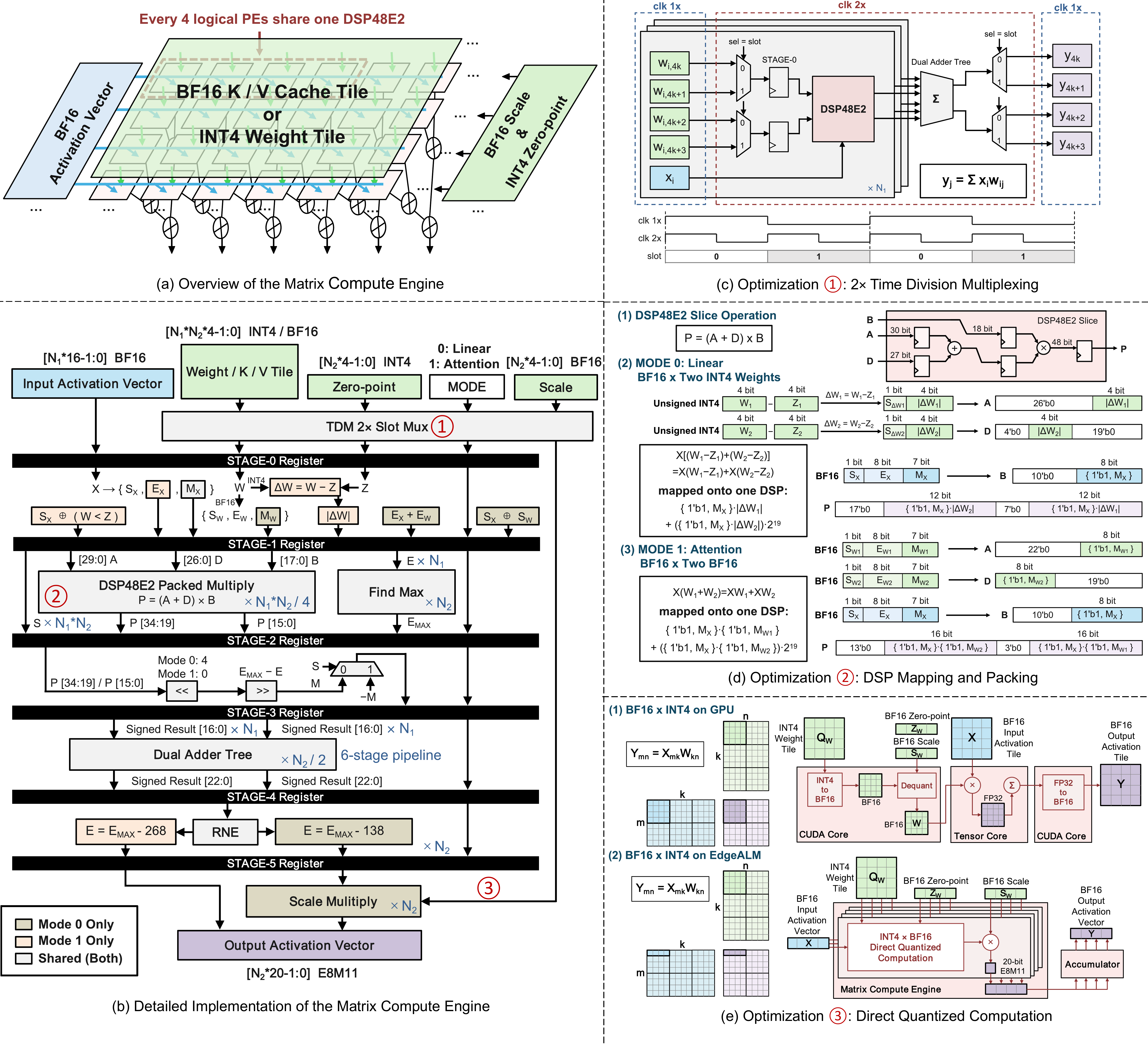"}
\caption{Implementation and optimization details of the Matrix Compute Engine.}
\label{fig:engine-microarch}
\end{figure*}

FPGA-based LLM accelerators commonly face memory-bound LLM decoding~\cite{flightllm,accllm}. 
However, EdgeLLM~\cite{edge_llm}'s grouped vector systolic array design places its 
decode-stage Linear operations at the Roofline ridge point.
With our more aggressive Matrix Compute Engine (MCE) design,
we also reach the same ridge point
even with a smaller compute budget, as shown in Fig.~\ref{fig:roofline}.
This is achieved as follows:
With the limited compute resources available on an FPGA, 
a conventional systolic array that updates only one row/column of weights 
per cycle may fail to saturate the required bandwidth. 
The MCE therefore updates the weights of the entire array every cycle, 
keeping pace with the weight supply rate demanded by LLM decoding.
Specifically, the engine is designed as a non-systolic, direct-wiring multicast array, 
as illustrated in Fig.~\ref{fig:engine-microarch}(a). 
In every cycle of LLM decoding, the array consumes one BF16 activation vector together 
with one complete weight tile whose size exactly matches the logical array 
(unlike the variable-sized programming-level tiles discussed in Sections II-C and V-A).
The activations are multicast along the rows through direct wiring, 
so that all PEs in the same row receive the same activation element in the same cycle; 
meanwhile, the entire weight tile is written into the weight registers of all PEs in a single cycle, 
i.e., the array switches to a completely new weight tile every cycle. 
No data is ever propagated between PEs: each PE locally multiplies one activation with one weight, 
and all the products in each column are reduced by a dedicated adder tree
at the bottom of each column, 
producing one partial sum per column per cycle. 
For less bandwidth-hungry workloads in an ALM,
a weight tile can also remain in the weight registers for multiple cycles.

We also apply several optimization techniques to improve our mixed-precision compute engine. 
In EdgeALM, the HBM memory subsystem introduced in Section~\ref{sec:hbm}
provides a bandwidth ceiling of 81.92~GB/s, forming the bandwidth roof, as shown in Fig.~\ref{fig:roofline}.
The peak compute of the MCE forms the INT4 compute roof. The two roofs 
intersect at a ridge point of 4 OP/B, precisely matching the arithmetic intensity of INT4-weight 
linear layers in LLM decoding.
In EdgeLLM's mixed-precision compute engine, each PE maps to 1 DSP48E2 in linear mode and 
3 DSP48E2s in attention mode. Its attention mode delivers only 25\% of the peak compute 
available in linear mode, as shown in Fig.~\ref{fig:roofline}. 
By contrast, our optimizations reduce the average usage to only 1/4 of a DSP48E2 
per PE in both modes, resulting in more efficient DSP use.
The peak compute ratio of attention mode to linear mode increases
from 25\% in EdgeLLM to 100\% in EdgeALM, a 4$\times$ improvement.
The following paragraphs detail the implementation of our MCE
and the optimization techniques it uses.

\subsubsection{Detailed Implementation of the Matrix Compute Engine}
Fig.~\ref{fig:engine-microarch}(b) shows the complete datapath of the Matrix Compute
Engine with an $N_1 \times N_2$ compute array.
Zero-point subtraction (INT4) or sign/exponent preprocessing (BF16) is
first performed, and the multiply-accumulate computation is then mapped onto DSPs.
The $N_1$ products in each column are aligned to the shared exponent given by Find\_Max,
assigned the appropriate signs, and reduced by the dual adder tree into two sums, which are
packed into the custom 20-bit E8M11 format for output.
In the linear mode, the results are further multiplied by the per-column dequantization
scale.

\subsubsection{DSP Mapping and Packing}
Details of the DSP mapping and packing optimization are presented
 in Fig.~\ref{fig:engine-microarch}(d).
Exploiting the bit-width headroom of the DSP48E2 multiplier, two weight mantissas are
packed into a single operand, so that one multiplication yields two products at once,
extracted from
the P output.
In the linear mode, the BF16 activation mantissa multiplies two INT4 weights with
their zero-points subtracted; in the attention mode, it multiplies two BF16 operands.
The per-DSP throughput is thus doubled.

\subsubsection{Time-Division Multiplexing}
We use 2$\times$ time-division multiplexing (TDM).
The weight, activation, and output paths run at clk1x, while the
compute pipeline runs at clk2x (Fig.~\ref{fig:engine-microarch}(c)). 
Within each clk1x cycle, the
pipeline processes two sets of weights, zero-points, and scales in
separate slots, sharing the same activation vector. Slot tags align
the results at the clk1x output boundary. 
Combining TDM with DSP packing allows four logical PEs to share one DSP48E2. 
All four results are emitted synchronously at the clk1x boundary.

\subsubsection{Direct Quantized Computation}
Fig.~\ref{fig:engine-microarch}(e) contrasts the BF16$\times$INT4
computation paths on a GPU and EdgeALM. In the GPU path, each INT4
weight tile must first be converted and dequantized to BF16 in CUDA
cores before being passed to Tensor Cores. EdgeALM instead sends the
INT4 weight tile, BF16 activation vector, zero-points, and BF16 scales
directly to the MCE. It computes the
INT4$\times$BF16 products and represents the resulting
partial sums in a custom 20-bit E8M11 format. The partial sums
are then accumulated to produce BF16 output activations. Since each
scale is shared by the weights in one output column, EdgeALM applies
it to the corresponding output element after computing the dot product,
rather than multiplying each weight by its scale. This reordering
reduces scale multiplications and avoids converting each INT4 weight
tile to BF16.

\subsection{Convolution Compute Engine}
\begin{figure}[t]
\centering
\includegraphics[width=\columnwidth]{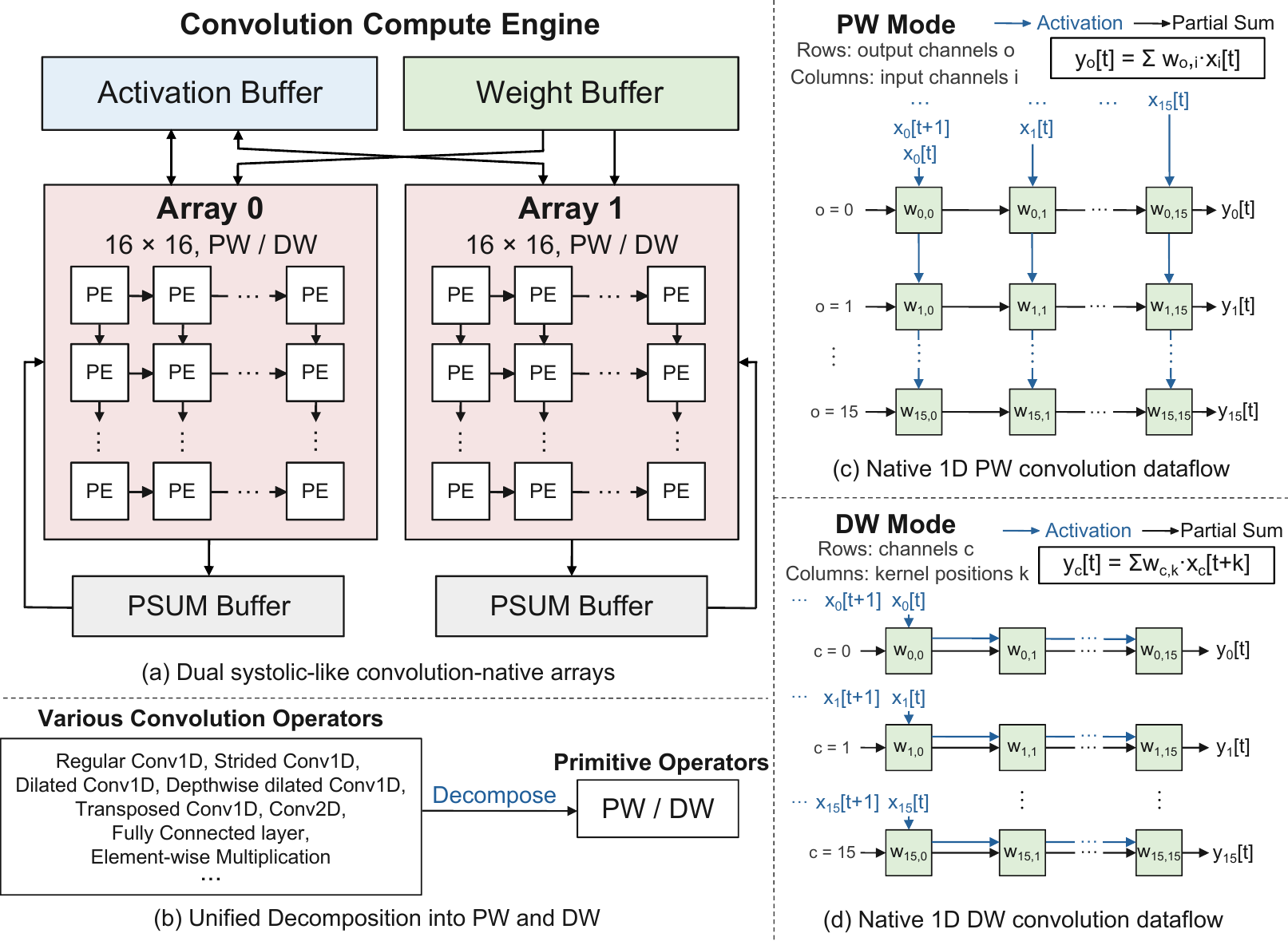}
\caption{An illustration of the Convolution Compute Engine.}
\label{fig:conv}
\end{figure}
The memory access demand of convolution is far lower than that of LLM decode.
Therefore, the Convolution Compute Engine (CCE) does not need the extreme design of the matrix path;
this allows it to adopt a systolic-array-style design, where data is forwarded between
adjacent PEs and the weights are updated column by column.
Precisely because of this low memory access demand, the CCE can work
simultaneously with the MCE while sharing the same HBM bandwidth, with limited
impact on the matrix path's throughput.

The CCE consists of two arrays, as shown in Fig.~\ref{fig:conv}.
Following the idea of the compute array in our previous work~\cite{u_sparta},
the arrays natively and efficiently support 1D pointwise (PW) convolution and 1D
depthwise (DW) convolution, and with these two as primitive operators, all other
convolutions are decomposed into PW or DW efficiently through various operator decomposition
techniques.
This is motivated by the observation
that the
convolutions in ALMs
almost all fall into the Conv1D category,
including regular, pointwise, depthwise, strided, dilated, and
transposed 1D convolutions, 
and can be efficiently decomposed into 1D PW or DW convolutions.
This enables efficient and broad support for the convolution workloads of diverse ALMs.
Such a convolution-native engine offers another benefit: it requires no Img2col
module.
As works~\cite{manasi_dwconv,spots} have shown, the Img2col method can lead to
inefficiency.
This is because Img2col transforms the activations rather than the weights, but the
convolution weights,
in speech applications in particular  
(e.g., the
depthwise convolution kernels), are often small or narrow, making it difficult to fully utilize
the entire array. Our convolution-native engine is free from these problems.

\section{ISA and Agent-Driven Recursive Optimization}
\label{sec:isa}
\begin{figure*}[!t]
\centering
\includegraphics[width=\textwidth]{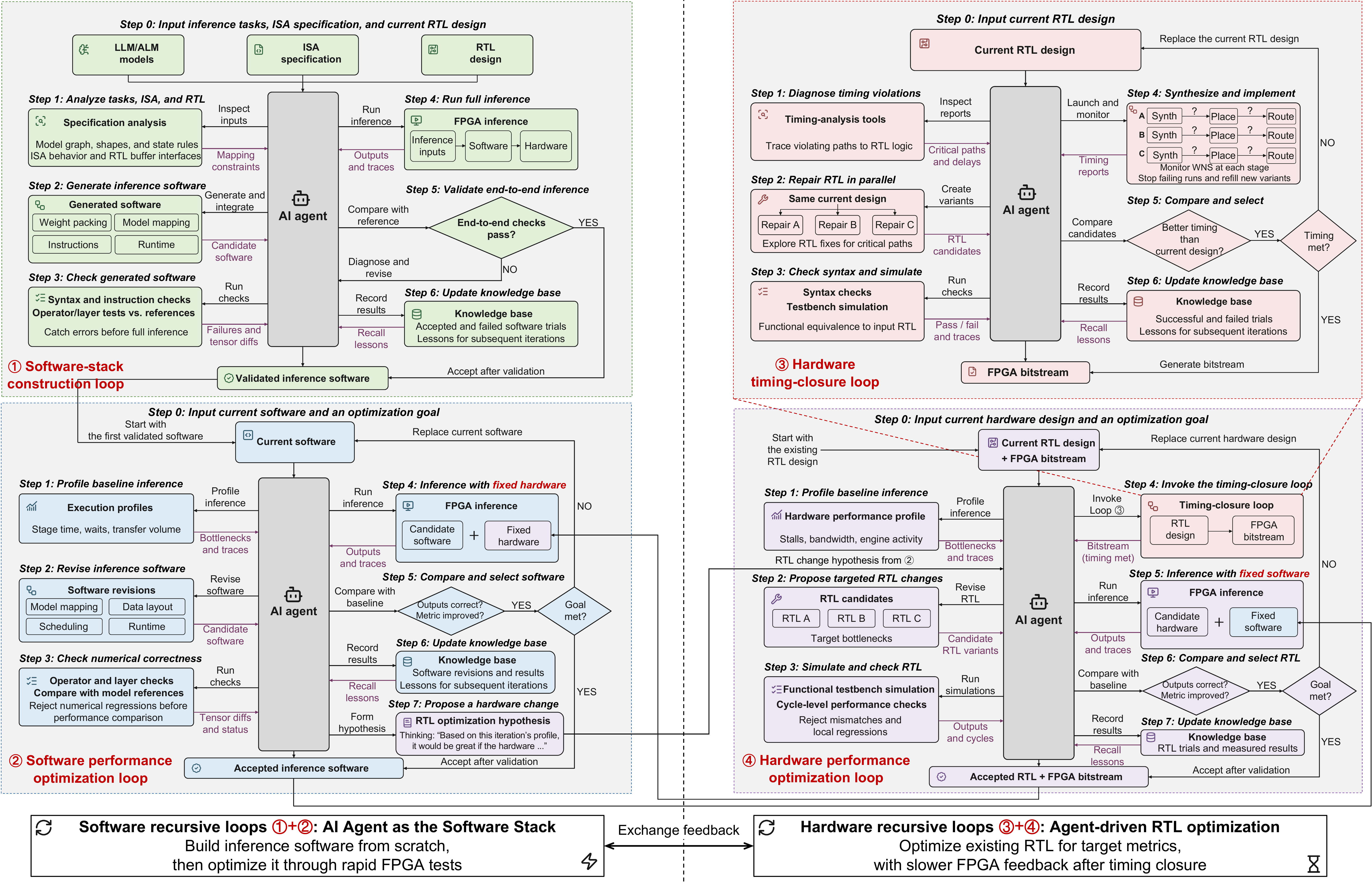}
\caption{Agent-driven recursive loops for developing and optimizing EdgeALM software and hardware.}
\label{fig:agent-stack}
\end{figure*}

\subsection{Extra-Coarse-Grained ISA}
As discussed in Section~II-C, on-chip buffer accesses and
execution sequences exhibit regularity that is often still
managed explicitly by software
when computation uses coarse-grained instructions.
To exploit this regularity in hardware, we propose an
\emph{Extra-Coarse-Grained ISA} for EdgeALM.
The ISA distinguishes two levels of data movement:
transfers between HBM and on-chip buffers, and 
transfers between on-chip buffers and
compute engines.
Transfers between HBM and on-chip buffers are \emph{explicit}
and \emph{decoupled} from computation.
DMA instructions specify HBM addresses, on-chip sources or
destinations, and transfer extents, and execute through the
Custom DMA Engine in streams independent of compute instructions.
Software thus controls data placement,
prefetching, and dependencies.
As shown in Fig.~\ref{fig:edgealm-architecture}, 
matrix compute, matrix DMA, convolution compute,
and convolution DMA form four independent logical instruction
streams. 
Both DMA streams share the
Custom DMA Engine. 

By contrast, data transfers between on-chip buffers and compute
engines are not exposed as separate instructions; each compute
instruction implicitly incorporates these transfers and
execution sequences, and couples
them with the computation.
This implicit coupling is what makes the ISA extra-coarse-grained.
Once software specifies the tile dimensions, buffer
locations, and reuse choices, a single compute command covers
the repeated buffer accesses, array operations, and result
updates needed to complete that computation.
This extra granularity reduces software complexity while keeping
hardware control simple
by exploiting
the regularity of on-chip accesses and iterations. Simple hardware 
mechanisms such as counters, address generators,
and finite-state control can drive these repeated steps with
modest control-logic overhead.
In this way, the design also amortizes instruction fetch, decode,
and issue across repeated steps and avoids storing lots of fully expanded
instruction sequences.
We also find empirically this ISA particularly well suited to our
\emph{AI Agent as the Software Stack} workflow.
We believe this suitability stems from the fact that such
coarse-grained instructions are semantically meaningful to an
AI agent, allowing it to reason directly about model mapping,
tiling, and reuse instead of generating and validating long
low-level instruction sequences.

The above exploits regularity within a tile.
At a higher level, a descriptor mechanism is introduced in
EdgeALM to further exploit potential regularity across tiles,
as shown in Fig.~\ref{fig:edgealm-architecture}.
Each descriptor specifies the compute and DMA instruction ranges
to execute, together with their HBM base addresses and initial
event state.
A sequence of descriptors can in turn be described by a
coarser-grained descriptor.
Tiles with the same execution pattern can therefore share
instruction sequences while accessing different data, without
regenerating or storing the sequences for each instance.
Within a single launch, the sequencer executes multiple
descriptors, avoiding a separate host launch for each descriptor
and reducing host launch overhead and host-induced idle gaps.

\subsection{Software recursive loops: AI Agent as the Software Stack}
In general, substantial engineering effort is required to 
build a software stack for a custom accelerator with a new ISA.
With AI agents, however, we see an opportunity to change
this situation.
We explore two software recursive loops,
collectively termed \emph{AI Agent as the Software Stack}.
Given the models to be deployed and the accelerator's ISA and RTL,
the agent acts as the whole software stack, directly generating
the software needed to run inference for those models on the hardware.
The workflow comprises
two software recursive loops: software-stack construction and
software performance optimization, as shown in Fig.~\ref{fig:agent-stack}.
In the construction loop, the agent generates candidate software
from scratch,
checks its syntax and instructions, and compares
outputs with model references. It 
then runs complete inference on
the FPGA and checks the outputs. Failures at either stage
guide the next revision. 
The performance-optimization loop starts with the first validated
implementation and an optimization goal.
With hardware fixed, the agent profiles inference,
identifies bottlenecks, and revises model mappings, data layouts,
schedules, or runtime behavior. It checks each candidate for numerical
correctness before measuring it on the FPGA, then compares it with the
current software. Validated
improvements become the next baseline. Since software revisions
require no FPGA rebuild, these loops obtain rapid on-board feedback.
Both loops record successful and failed trials and their lessons
in their respective knowledge bases to guide later iterations.
Software profiling may also suggest RTL changes
for the hardware loops.
Humans are in the loop to set optimization goals, 
monitor progress, and decide whether the proposed workflow needs improvement.

\subsection{Hardware recursive loops: Agent-driven RTL optimization}
Unlike the software recursive loops, which build the stack
entirely from scratch, EdgeALM's hardware design is already
clearly defined by its Extra-Coarse-Grained ISA, CM Disaggregation architecture,
and microarchitectural details as described above.
This well-specified design made it straightforward to
implement an initial RTL design.
The hardware recursive loops therefore begin with that existing RTL design.
The hardware side comprises
a timing-closure loop and a performance-optimization loop (Fig.~\ref{fig:agent-stack}).
The timing-closure loop generates a bitstream from the input RTL
design as a stage of the performance-optimization loop, addressing
the routing congestion and timing-closure difficulties of large designs.
In the timing-closure loop, the agent
analyzes critical paths, explores functionally equivalent
RTL changes to improve timing, and runs synthesis, placement, and routing.
It uses the timing
results to select the next parent design and repeats the process
until timing closes and an FPGA bitstream can be generated.
The performance-optimization loop begins with the current RTL design
and bitstream, an optimization goal, and fixed inference software.
The agent profiles FPGA inference to identify stalls and bandwidth
limits and assess compute-engine utilization.
It proposes RTL changes targeted at
these bottlenecks and checks their behavior and cycle-level performance
in simulation. 
The agent can repeat the relatively fast steps 1--3
to accumulate promising RTL changes for each candidate, and then
invoke the slower timing-closure loop to produce a bitstream.
It then runs the same inference software
on each candidate, checks the outputs, and compares the target metric 
with that of the current design. Validated improvements become the starting
point for subsequent iterations.
Unlike software revisions, RTL changes require new FPGA
builds and may involve multiple timing-closure iterations, making
hardware feedback slower.

\begin{figure}[!t]
\begin{minipage}{\columnwidth}
\centering
  \includegraphics[width=\columnwidth]{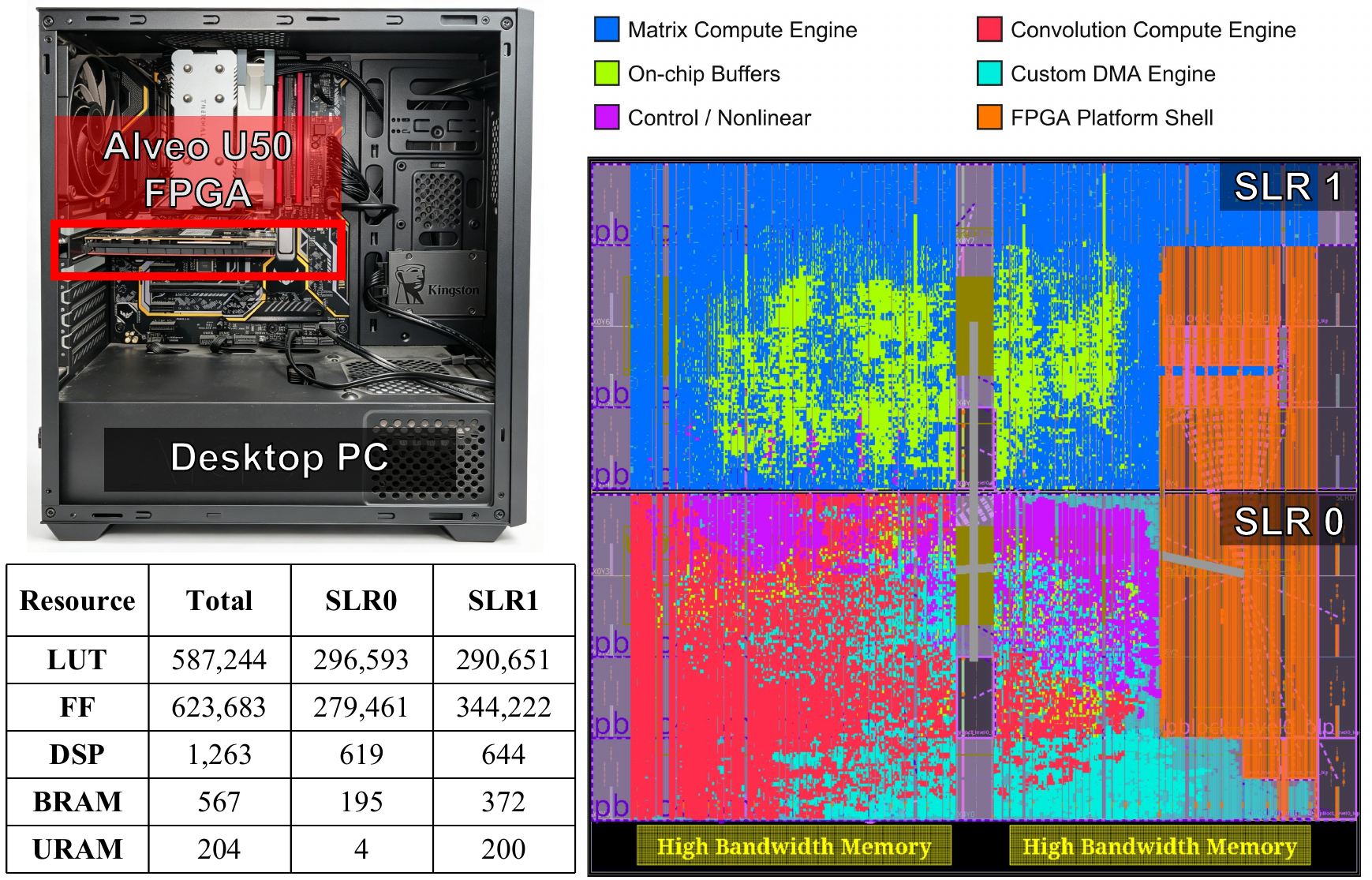}
  \caption{Hardware platform, resource consumption, and implementation layout of EdgeALM on the Alveo U50 FPGA.}
  \label{fig:u50}
\end{minipage}
\par\vspace{\floatsep}
\begin{minipage}{\columnwidth}
\makeatletter\def\@captype{table}\makeatother
\footnotesize
\centering
  \caption{Compression effects on ALM quality and weight storage.}
  \label{tab:alm-compression}
  \begingroup
  \setlength{\tabcolsep}{3pt}
  \renewcommand{\arraystretch}{1.0}
  \renewcommand{\tabularxcolumn}[1]{m{#1}}
  \setlength{\aboverulesep}{0.25ex}
  \setlength{\belowrulesep}{0.35ex}
  \newcommand{\compressionpair}[2]{\parbox[c][0.24in][c]{\linewidth}{\centering\setlength{\baselineskip}{8.5pt}#1\\#2}}
  \newcommand{\compressionheading}[1]{\parbox[c][0.24in][c]{\linewidth}{\centering #1}}
  \begin{tabularx}{\columnwidth}{>{\centering\arraybackslash\bfseries}m{0.99in}>{\centering\arraybackslash}m{0.60in}>{\centering\arraybackslash}X>{\centering\arraybackslash}m{0.74in}}
    \toprule
    Metric
      & \textbf{A2T score}
      & \textbf{WER / CER (\%)}
      & \textbf{Weights (GB)} \\
    \midrule
    \compressionheading{Original}
      & \compressionpair{69.90}{41.61}
      & \compressionpair{5.18 / 3.07}{2.91 / 2.15}
      & \compressionpair{25.14}{9.68} \\
    \compressionheading{Vocabulary pruning}
      & \compressionpair{69.90}{41.61}
      & \compressionpair{5.18 / 3.07}{2.91 / 2.15}
      & \compressionpair{23.94}{9.41} \\
    \compressionheading{Quantization}
      & \compressionpair{66.68}{39.95}
      & \compressionpair{6.37 / 4.18}{3.61 / 3.05}
      & \compressionpair{8.56}{3.02} \\
    \midrule
    \compressionheading{Combined}
      & \compressionpair{66.68}{39.95}
      & \compressionpair{6.37 / 4.18}{3.61 / 3.05}
      & \compressionpair{7.35}{2.75} \\
    \bottomrule
  \end{tabularx}
  \endgroup
  \par\vspace{2pt}
  \begin{minipage}{\columnwidth}
    \scriptsize\raggedright
    In each cell, the upper and lower values correspond to Kimi-Audio
    and LLaMA-Omni2 -3B, respectively.
    A2T denotes the audio-to-text chat task on the Alpaca Eval dataset, 
    while WER and CER denote the word and character error rates between the generated speech 
    and its corresponding text in the audio-to-audio task.
    A2T measures the quality of the generated text, 
    while WER/CER measures the quality of the generated speech.
  \end{minipage}
\end{minipage}
\end{figure}

\begin{figure*}[!t]
\begin{minipage}{\textwidth}
\centering
  \includegraphics[width=\textwidth]{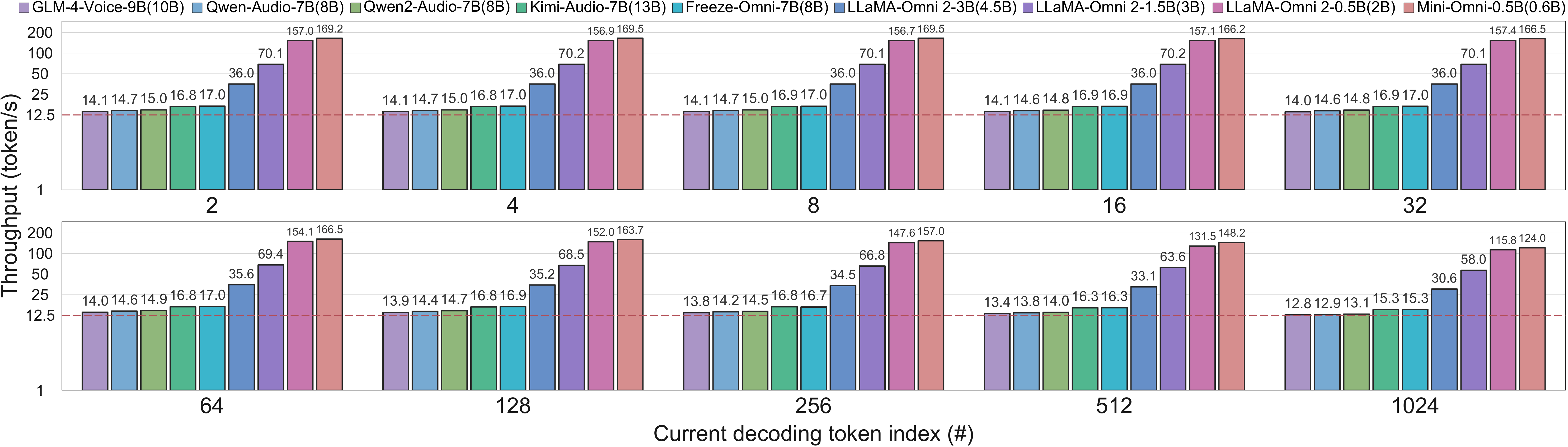}
  \caption{Audio-to-text chat decode throughput of diverse ALMs on EdgeALM. Model-xB(yB): x = LLM backbone parameters; y = total ALM parameters.}
  \label{fig:decode}
\end{minipage}
\par\vspace{\floatsep}
\begin{minipage}{\textwidth}
\makeatletter\def\@captype{table}\makeatother
\footnotesize
\caption{Streaming audio-to-audio inference on Jetson AGX
Orin and EdgeALM. Real-time requires
$\mathrm{RTF_S}\leq 1$ after the first chunk.}
\label{tab:a2a_streaming}
\centering
\setlength{\tabcolsep}{7pt}
\setlength{\extrarowheight}{1.5pt}
\setlength{\aboverulesep}{0pt}
\setlength{\belowrulesep}{0pt}
\begin{tabular}{ccccccc}
\specialrule{0.6pt}{0pt}{0pt}
Platform & Model & Duplex & $\mathrm{RTF_S}$ & Real-time? & First-response
latency (s) & Power (W) \\
\hline
\multirow{2}{*}[-1.2pt]{NVIDIA Jetson AGX Orin 64GB} & Freeze-Omni & Full-duplex & 2.82 & \ding{55} & 3.96 & 42.09 \\
 & Mini-Omni   & Half-duplex & 1.16 & \ding{55} & 3.44 & 34.96 \\
\midrule[0.1pt]
\multirow{2}{*}[-1pt]{EdgeALM} & Freeze-Omni & Full-duplex & 0.96 & \ding{51} & 2.97 & 26.84 \\
 & Mini-Omni   & Half-duplex & 0.58 & \ding{51} & 2.21 & 26.37 \\
\specialrule{0.6pt}{0pt}{0pt} 
\end{tabular}
\par\vspace{2pt}
\begin{minipage}{\textwidth}
\scriptsize\raggedright
For $N$ output chunks, let $t_i$ and $d_i$ be the arrival time and playback duration of chunk $i$,
and $q_i$ the buffered playable duration immediately after its arrival. Playback starts with the first chunk:
$q_1=d_1$. With $g_i=t_i-t_{i-1}$, $q_i=d_i+\max(0,q_{i-1}-g_i)$ and
$\mathrm{RTF_S}=\max_{2\leq i\leq N}(g_i/q_{i-1})$. 
Under this definition, sustaining real-time streaming audio-to-audio
inference after the first chunk arrives is equivalent to
$\mathrm{RTF_S}\leq 1$.

\end{minipage}
\end{minipage}
\par\vspace{\floatsep}
\begin{minipage}{\textwidth}
\makeatletter\def\@captype{table}\makeatother
\footnotesize
\centering
\caption{Examples of ASR, text-only LLM, and TTS models and their inference performance on EdgeALM.}
\label{tab:other-models}
\setlength{\tabcolsep}{3.5pt}
\renewcommand{\arraystretch}{1.0}
\setlength{\extrarowheight}{1.5pt}
\setlength{\arrayrulewidth}{0.35pt}
\setlength{\aboverulesep}{0pt}
\setlength{\belowrulesep}{0pt}
\begin{tabular}{cc>{\hspace{6pt}}c<{\hspace{6pt}}@{\hspace{7pt}}|@{\hspace{7pt}}ccc@{\hspace{7pt}}|@{\hspace{7pt}}cc>{\hspace{6pt}}c<{\hspace{6pt}}}
\toprule
& Model & RTF &
& Model & Decode speed (tokens/s) &
& Model & RTF \\ 
\midrule
\multirow{4}{*}[-1pt]{\textbf{ASR}} & Moonshine-tiny & 0.071 &
\multirow{4}{*}[-1pt]{\textbf{LLM}} & Qwen2-0.5B & 163.8 &
\multirow{4}{*}[-1pt]{\textbf{TTS}} & YourTTS & 0.387 \\
& Whisper-tiny   & 0.095 & & Qwen2.5-3B & 35.4 & & MMS-TTS & 0.579 \\
& Whisper-base     & 0.146 & & Qwen2.5-7B & 16.8 & & JETS & 0.782 \\
& Qwen3-ASR-0.6B   & 0.286 & & GLM-4-9B & 14.7 & & VITS & 0.822 \\
\bottomrule
\end{tabular}
\par\vspace{2pt}
\begin{minipage}{\textwidth}
\scriptsize\raggedright
For ASR and TTS offline inference, we report RTF rather than the streaming metric $\mathrm{RTF_S}$.
RTF is the time from complete input availability to the final transcript (ASR) or complete waveform (TTS),
divided by the input (ASR) or generated (TTS) audio duration.
\end{minipage}
\end{minipage}
\par\vspace{\floatsep}
\noindent
\begin{minipage}[t]{\columnwidth}
\vspace{0pt}
\centering
  \includegraphics[width=\columnwidth]{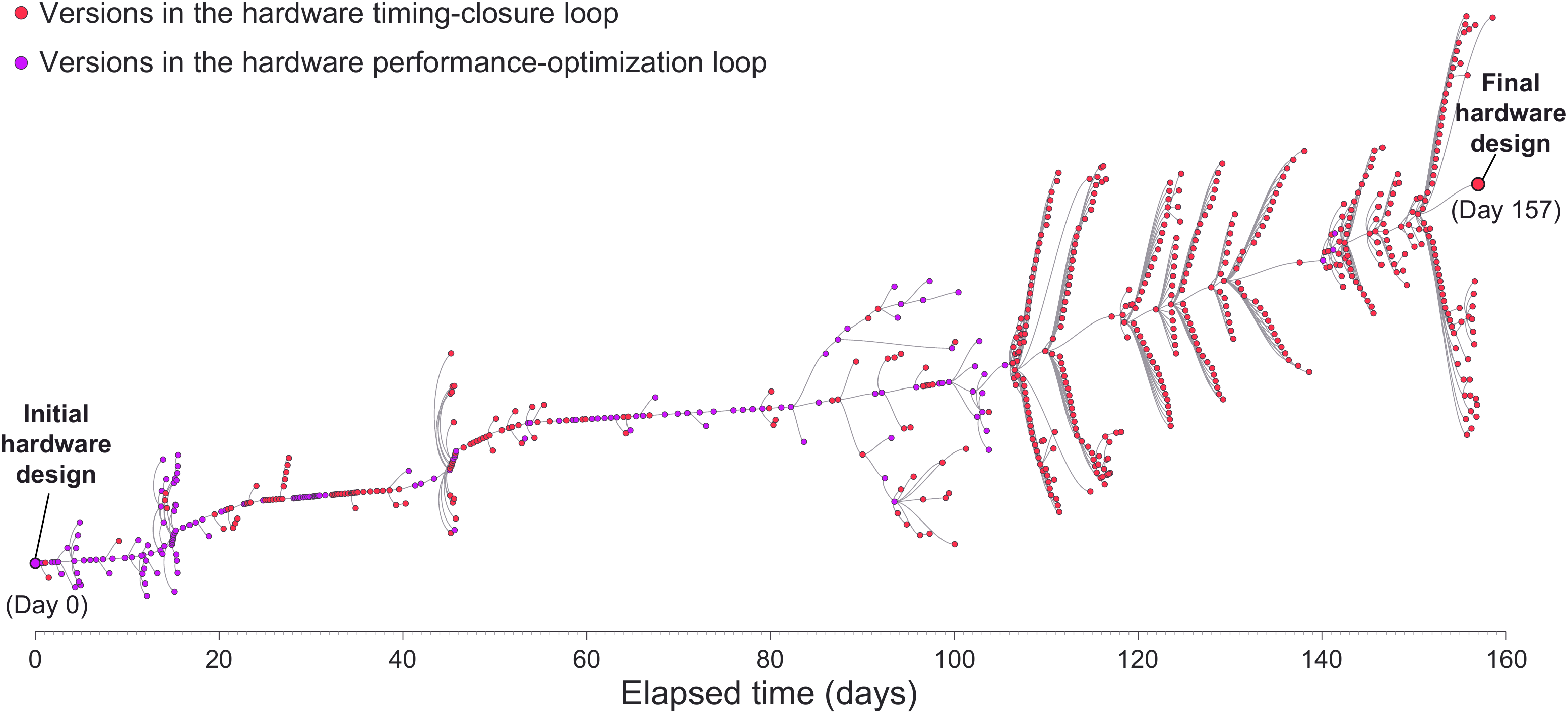}
  \caption{Evolution of EdgeALM through the hardware recursive
  loops. Each node represents a hardware design version, and each edge
  connects a parent version to a derived version.}
  \label{fig:hardwareloop}
\end{minipage}
\hfill
\begin{minipage}[t]{\columnwidth}
\vspace{0pt}
\centering
  \includegraphics[width=\columnwidth]{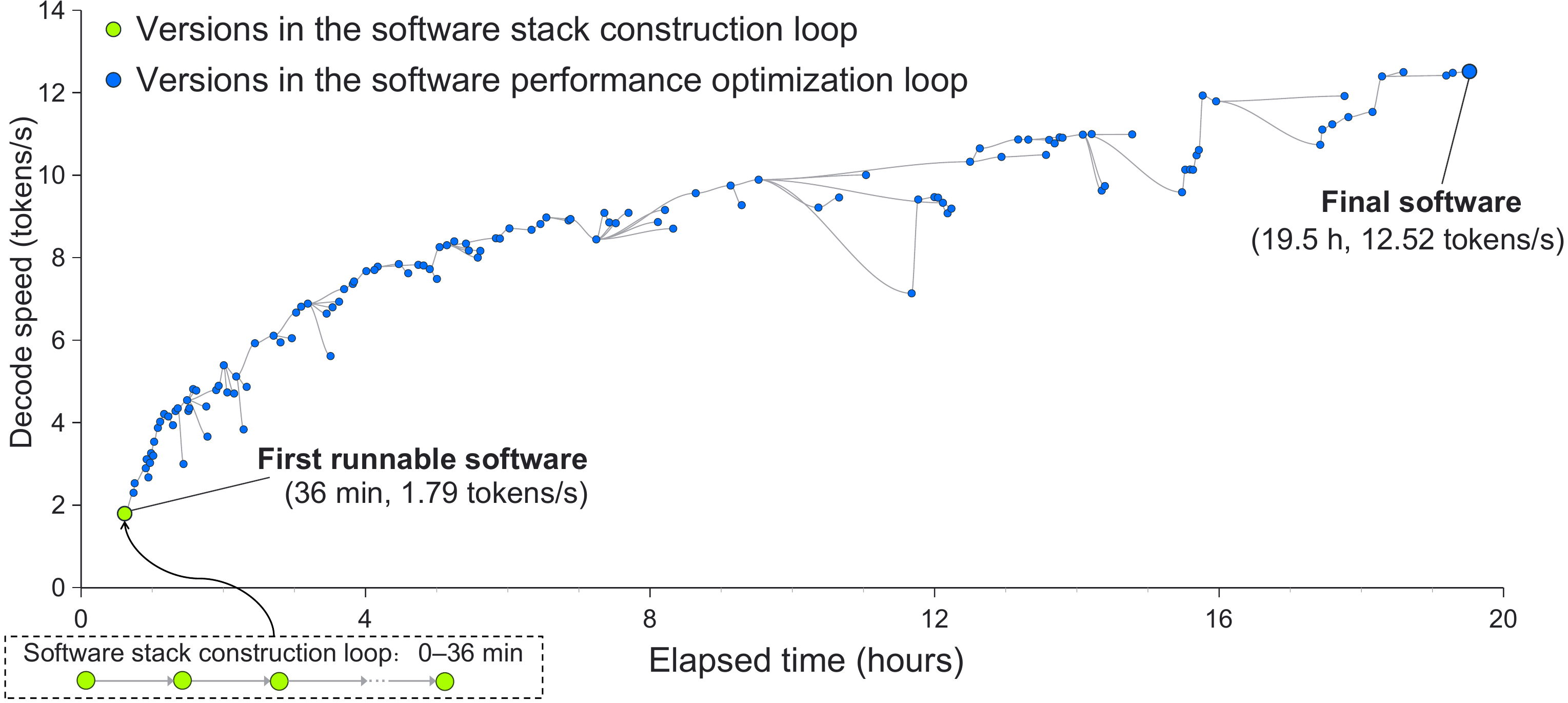}
  \caption{Evolution of software for audio-to-text inference with GLM-4-Voice
  on EdgeALM through the software recursive loops.
  Each node represents a software version, and each edge connects
  a parent version to a derived version.}
  \label{fig:softwareloop}
\end{minipage}
\end{figure*}

\section{Evaluation}
\label{sec:evaluation}
\subsection{Experimental Setup}
The proposed EdgeALM accelerator has been implemented on an Alveo U50 FPGA
using Vivado and Vitis 2022.2. 
The Alveo U50 FPGA is installed in an ordinary desktop PC
equipped with an Intel Core i7-8700 CPU
and cooled by a custom fan.
The fan keeps the FPGA temperature below 60°C, enabling stable operation.
The design is primarily clocked at 160 MHz, including the CCE, 
the MCE's clk2x domain, 
the Custom DMA Engine, all the buffers and control logic.
The HBM platform clock runs at 
450 MHz.

\subsection{Model Compression and FPGA Resource Utilization}
Table~\ref{tab:alm-compression} presents the experimental results
for the compression techniques described in
Section~\ref{sec:quantization}.
Vocabulary pruning preserves the reported A2T scores and WER/CER,
whereas quantization reduces the A2T scores by 3.22 and 1.66 points,
respectively, with WER and CER increasing by at most 1.19 percentage
points on the evaluated samples.
The combination of quantization and vocabulary pruning 
reduces the weight footprint from
25.14 to 7.35~GB for Kimi-Audio and from 9.68 to 2.75~GB for
LLaMA-Omni2-3B, corresponding to reductions of 70.8\% and 71.6\%,
respectively.
These results demonstrate substantial storage savings with limited
degradation in text response quality and speech content fidelity.

Fig.~\ref{fig:u50} shows the Alveo U50 installed in a desktop PC,
the routed EdgeALM layout, and the resource consumption. 
The MCE is concentrated
in SLR1, whereas the CCE is placed in SLR0.
Through iterations of the hardware recursive loops, the final design
achieves nearly balanced LUT usage across the two SLRs,
helping distribute placement pressure and ease timing closure.
It was found through the hardware recursive loops that
moving most of the matrix-side Nonlinear Unit from its planned location
in SLR1 (Fig.~\ref{fig:edgealm-architecture}) to SLR0 improved resource
balance, despite increased SLL usage. Meanwhile, most of the
URAMs reside in SLR1, consistent with the matrix-side buffer
requirements shown in Fig.~\ref{fig:edgealm-architecture}.

\subsection{Recursive Optimization Results}
Fig.~\ref{fig:hardwareloop} provides an overview of EdgeALM's
evolution through the hardware recursive loops.
These loops were run continuously on our development server with 256~GB of memory.
The process started with a basic implementation of our well-defined
ISA, architecture, and microarchitecture, which served as the initial
hardware design. The hardware recursive loops then evolved this
implementation into the final hardware design over 158 days,
with 705 design versions in total.
Most early versions were
developed in the hardware performance-optimization loop,
because the design was still relatively small and had
not yet been pushed to its performance limits.
As successive optimizations increased resource demand and pushed
performance further, placement and routing became increasingly
constrained, making timing closure more difficult.
Therefore, versions developed within the timing-closure loop
dominated the later stages of the evolution.
Fig.~\ref{fig:softwareloop} shows the evolution of software
on EdgeALM through the software recursive loops.
The software stack construction loop produced the first runnable
software from scratch in 36 minutes, achieving a decode throughput
of 1.79 tokens/s. The software performance-optimization loop then
refined this implementation with a human-specified
performance goal of achieving a decode speed above 12.5~tokens/s.
It ultimately reached 12.52 tokens/s
at 19.5 hours, after a total of 127 versions.
One important optimization that the agent identified and implemented
through these iterations was the consolidation of all model
inference operations into a single FPGA launch,
reducing intermediate writebacks to HBM and transfers to the host.
The ability to build software from scratch and efficiently optimize
it
may be facilitated by our
proposed Extra-Coarse-Grained ISA and descriptor mechanism.

\subsection{Performance Comparison}
\begin{table}[!t]
\centering
\newcommand{\comparisonmetricwidth}{0.54in}
\newcommand{\comparisonheadfont}{\fontsize{6pt}{7pt}\selectfont}
\newcommand{\comparisoncorner}[1]{%
  \multicolumn{1}{|@{}c@{}|}{$\vcenter{\hbox{%
    \begin{tikzpicture}[x={\dimexpr\comparisonmetricwidth+2\tabcolsep\relax},y=#1,
        every node/.style={inner sep=0pt,outer sep=0pt,font=\comparisonheadfont\bfseries}]
      \path[use as bounding box] (0,0) rectangle (1,1);
      \begin{scope}[overlay]
        \draw[line width=\arrayrulewidth] (0,1) -- (1,0);
        \node at ({1/3},{1/3}) {Metric};
        \node at ({2/3},{2/3}) {Work};
      \end{scope}
    \end{tikzpicture}}}$}}
\begin{minipage}{\columnwidth}
\centering
\footnotesize
\caption{Comparison of ASR Accuracy Across Audio Accelerators}
\label{tab:asr_cas_comparison}
\scriptsize
\renewcommand{\comparisonmetricwidth}{0.65in}
\newcommand{\asrcomparisonvaluewidth}{0.503in}
\newcommand{\asrcomparisonlabel}[1]{\parbox[c][0.22in][c]{\comparisonmetricwidth}{\centering\textbf{#1}}}
\newcommand{\asrcomparisonheading}[1]{\parbox[c][0.22in][c]{\asrcomparisonvaluewidth}{\centering\comparisonheadfont #1}}
\setlength{\tabcolsep}{0.9pt}
\begin{tabular}{|c|c|c|c|c|c|}
\hline
\comparisoncorner{0.28in}
  & \asrcomparisonheading{\textbf{TCAS-II'22}\\\cite{liou_attention_asr}}
  & \asrcomparisonheading{\textbf{TCAS-I'24}\\\cite{yang_rsnn_asr}}
  & \asrcomparisonheading{\textbf{TCAS-I'26}\\\cite{u_sparta}}
  & \asrcomparisonheading{\textbf{JSSC'23}\\\cite{tambe_sm6}}
  & \asrcomparisonheading{\textbf{This work}} \\
\hline
\asrcomparisonlabel{Model}
  & \parbox[c][0.25in][c]{\asrcomparisonvaluewidth}{\centering Attention\\GRU}
  & RSNN & QuartzNet & LAS
  & \parbox[c][0.25in][c]{\asrcomparisonvaluewidth}{\centering
    \fontsize{6.4pt}{7.5pt}\selectfont
    \resizebox{0.44in}{!}{Qwen3-ASR-}\\0.6B} \\
\hline
\asrcomparisonlabel{Size (MB)}
  & 0.21 & 0.10 & 2.49 & 3.5 & 955.9 \\
\hline
\asrcomparisonlabel{Params. (M)}
  & 0.34 & 0.202 & 1.24 & 3.5 & 938.0 \\
\hline
\asrcomparisonlabel{{\fontsize{6.3pt}{7pt}\selectfont WER/PER (\%)}}
  & 22.80\textsuperscript{a} & 22.6\textsuperscript{a}
  & 12.40\textsuperscript{b} & 9.40\textsuperscript{b} & 3.39\textsuperscript{b} \\
\hline
\end{tabular}
\par\vspace{2pt}
{\scriptsize\raggedright
\textsuperscript{a}\,Phoneme Error Rate on TIMIT. \textsuperscript{b}\,Word Error Rate on LibriSpeech.
}
\end{minipage}
\par\vspace{\floatsep}
\begin{minipage}{\columnwidth}
\centering
\footnotesize
\caption{Comparison of the SOTA FPGA-Based LLM Accelerators}
\label{tab:fpga-comparison}
\scriptsize
\newcommand{\comparisonlabel}[2]{\parbox[c][#1][c]{\comparisonmetricwidth}{\centering\bfseries\shortstack[c]{#2}}}
\newcommand{\comparisonvalue}[2]{\parbox[c][#1][c]{0.525in}{\centering\shortstack[c]{#2}}}
\newcommand{\comparisonheading}[2]{{\comparisonheadfont\comparisonvalue{0.30in}{\textbf{#1}\\#2}}}
\newcommand{\comparisonmodel}[1]{{\fontsize{5.6pt}{6.5pt}\selectfont #1}}
\setlength{\tabcolsep}{0.9pt}
\begin{tabular}{|c|c|c|c|c|c|}
\hline
\comparisoncorner{0.30in}
  & \comparisonheading{FPGA'24}{\textbf{FlightLLM}~\cite{flightllm}}
  & \comparisonheading{TCAS-I'25}{\textbf{EdgeLLM}~\cite{edge_llm}}
  & \comparisonheading{TCAS-I'25}{\textbf{CLAT}~\cite{clat}}
  & \comparisonheading{TVLSI'26}{\textbf{AccLLM}~\cite{accllm}}
  & \comparisonheading{This work}{\textbf{EdgeALM}} \\
\hline
\comparisonlabel{0.28in}{Modality}
  & \comparisonvalue{0.28in}{Text-only}
  & \comparisonvalue{0.28in}{Text-only}
  & \comparisonvalue{0.28in}{Text-only}
  & \comparisonvalue{0.28in}{Text-only}
  & \comparisonvalue{0.28in}{Text \&\\Audio} \\
\hline
\comparisonlabel{0.33in}{Model}
  & \comparisonvalue{0.33in}{\comparisonmodel{OPT-6.7B}\\\comparisonmodel{LLaMA2-7B}}
  & \comparisonvalue{0.33in}{\comparisonmodel{ChatGLM2-6B}\\\comparisonmodel{Qwen2-7B}}
  & \comparisonvalue{0.33in}{\comparisonmodel{Llama2-7B}\\\comparisonmodel{OPT-6.7B}\\\comparisonmodel{Llama3-8B}}
  & \comparisonvalue{0.33in}{\comparisonmodel{LLaMA2-7B}}
  & \comparisonvalue{0.33in}{\comparisonmodel{Over 20}\\\comparisonmodel{models}} \\
\hline
\comparisonlabel{0.24in}{Max Size}
  & \comparisonvalue{0.24in}{7B}
  & \comparisonvalue{0.24in}{7B}
  & \comparisonvalue{0.24in}{8B}
  & \comparisonvalue{0.24in}{7B}
  & \comparisonvalue{0.24in}{13B} \\
\hline
\comparisonlabel{0.28in}{Platform}
  & \comparisonvalue{0.28in}{Alveo\\U280}
  & \comparisonvalue{0.28in}{Xilinx\\VCU128}
  & \comparisonvalue{0.28in}{Alveo\\U280}
  & \comparisonvalue{0.28in}{Alveo\\U280}
  & \comparisonvalue{0.28in}{Alveo\\U50} \\
\hline
\comparisonlabel{0.28in}{Memory}
  & \comparisonvalue{0.28in}{8GB HBM\\32GB DDR}
  & \comparisonvalue{0.28in}{8GB HBM\\32GB DDR}
  & \comparisonvalue{0.28in}{8GB HBM\\32GB DDR}
  & \comparisonvalue{0.28in}{8GB HBM\\32GB DDR}
  & \comparisonvalue{0.28in}{8GB HBM} \\
\hline
\comparisonlabel{0.24in}{Price}
  & \comparisonvalue{0.24in}{$\sim$\$8K}
  & \comparisonvalue{0.24in}{$\sim$\$10K}
  & \comparisonvalue{0.24in}{$\sim$\$8K}
  & \comparisonvalue{0.24in}{$\sim$\$8K}
  & \comparisonvalue{0.24in}{$\sim$\$3K} \\
\hline
\comparisonlabel{0.30in}{Bandwidth\\Utilization}
  & \comparisonvalue{0.30in}{65.9\%}
  & \comparisonvalue{0.30in}{$\sim$75\%}
  & \comparisonvalue{0.30in}{N/A}
  & \comparisonvalue{0.30in}{N/A}
  & \comparisonvalue{0.30in}{94\%} \\
\hline
\comparisonlabel{0.22in}{Power (W)}
  & \comparisonvalue{0.22in}{45}
  & \comparisonvalue{0.22in}{56.86}
  & \comparisonvalue{0.22in}{N/A}
  & \comparisonvalue{0.22in}{33}
  & \comparisonvalue{0.22in}{26} \\
\hline
\end{tabular}
\end{minipage}
\end{table}

Fig.~\ref{fig:decode} shows the audio-to-text chat decode speed
of different ALMs on EdgeALM. 
Each model's inference software was optimized through the
software recursive loops to improve decode throughput and bandwidth
utilization.
Among them, Mini-Omni reaches a peak of
169.5 tokens/s. 
The horizontal red line is a
threshold reflecting the fact that ALMs typically generate output speech
at 12.5 frames/s or higher, corresponding to at most 80 ms/frame. 
Sustaining at least 12.5 tokens/s in audio-to-text inference is therefore
a necessary condition for real-time streaming audio-to-audio
inference.
All models in
Fig.~\ref{fig:decode} exceed this rate threshold.
By contrast, embedded-FPGA accelerators~\cite{pushlimit,hummingbird}
report only about 5--9 tokens/s even for text-only LLMs, making them
unsuitable for real-time ALM inference,
as discussed in Section~II-B.
Across these tests, the measured power consumption averaged 26 W,
and bandwidth utilization averaged 94\% of EdgeALM's bandwidth ceiling.
The power is measured using the Xilinx tool xbutil.

We then evaluated the performance of real-time streaming audio-to-audio inference,
as presented in Table~\ref{tab:a2a_streaming}.
In this mode, output speech is generated and delivered to the
audio player in successive chunks.
We use a real-time factor $\mathrm{RTF_S}$ to assess whether real-time streaming
inference can be sustained.
It compares EdgeALM with an NVIDIA Jetson AGX Orin
on streaming audio-to-audio inference. EdgeALM sustains real-time
inference for both Freeze-Omni and Mini-Omni, whereas the Orin does not.
EdgeALM also achieves lower first-response latency
and consumes less power.
In addition to ALMs, EdgeALM supports standalone ASR, text-only LLM,
and TTS models, with their decent inference performance reported in
Table~\ref{tab:other-models}. 
These models can also be freely combined 
to form an ASR--LLM--TTS cascade system on EdgeALM.
Since EdgeALM supports recent billion-scale ASR models, our ASR accuracy
is higher than that of prior accelerators
(Table~\ref{tab:asr_cas_comparison}).

Table~\ref{tab:fpga-comparison} compares EdgeALM with four representative
state-of-the-art HBM-based FPGA accelerators for LLM inference.
Whereas the compared designs are evaluated on text-only models of up to 8B parameters,
EdgeALM supports a range of multimodal text-and-audio models of up to 13B parameters
and achieves 94\% of its bandwidth ceiling on a lower-cost Alveo U50
with an average board power of 26~W.


\section{Conclusion}
This paper presents EdgeALM, an efficient and unified FPGA
accelerator for ALMs through full-stack hardware-software co-design
and agent-driven recursive optimization. Its design combines
end-to-end ALM compression, Conv-Matmul Disaggregation with two
dedicated compute engines for matrix and convolution workloads, an
Extra-Coarse-Grained ISA, and software and hardware recursive loops.
Implemented on a single Alveo U50 installed in an ordinary desktop
PC, EdgeALM demonstrates inference across more than 20 models,
including ALMs with up to 13B parameters within 8~GB of HBM.
For ALMs, it achieves high-throughput audio-to-text inference and
sustains real-time streaming audio-to-audio inference.
It also supports standalone and cascaded ASR, text-only LLM, and TTS
inference.
Future work will explore an ASIC implementation
and extend EdgeALM's full-stack
co-design and recursive optimization
methodology to address the distinct workload characteristics
of more modalities.

\bibliographystyle{IEEEtran}
\bibliography{IEEEabrv,venue_abbr,my_ref}

\begingroup
\makeatletter
\renewcommand*{\@IEEEBIOskipN}{2\baselineskip}
\makeatother

\begin{IEEEbiography}[{\includegraphics[width=1in,height=1.25in,clip,keepaspectratio]{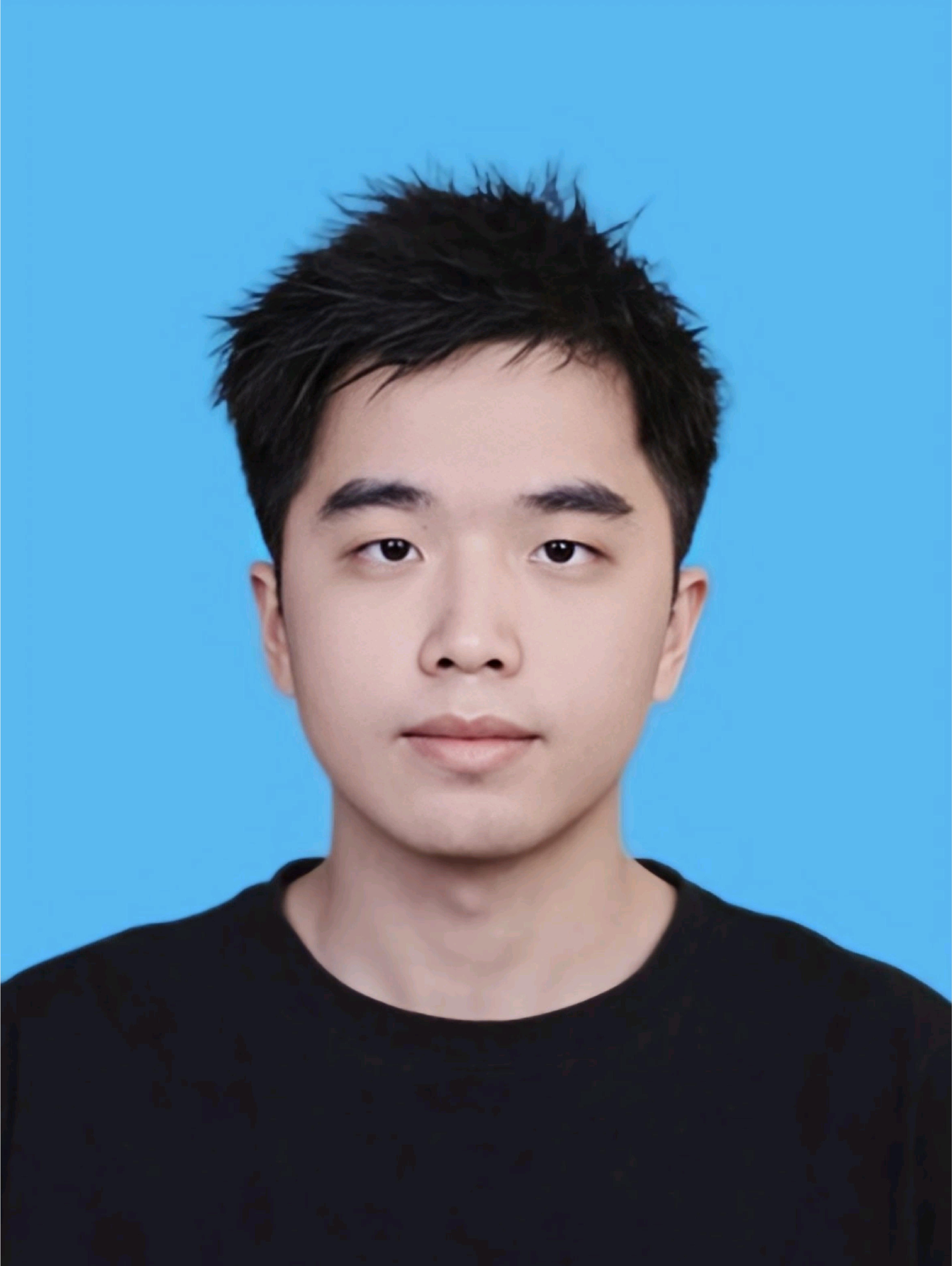}}]{Chenglin Yang}
received the B.S. degree in microelectronics science and engineering from
Sun Yat-sen University, Guangzhou, China, in 2023. He is currently pursuing
the Ph.D. degree with the School of Integrated Circuits, Sun Yat-sen University,
Shenzhen, China. His current research interests include inference acceleration of large language models (LLMs) 
for edge deployment, hardware–software co-design, speech, multimodal, and embodied AI, as well as AI for chip design.
\end{IEEEbiography}

\begin{IEEEbiography}[{\includegraphics[width=1in,height=1.25in,clip,keepaspectratio]{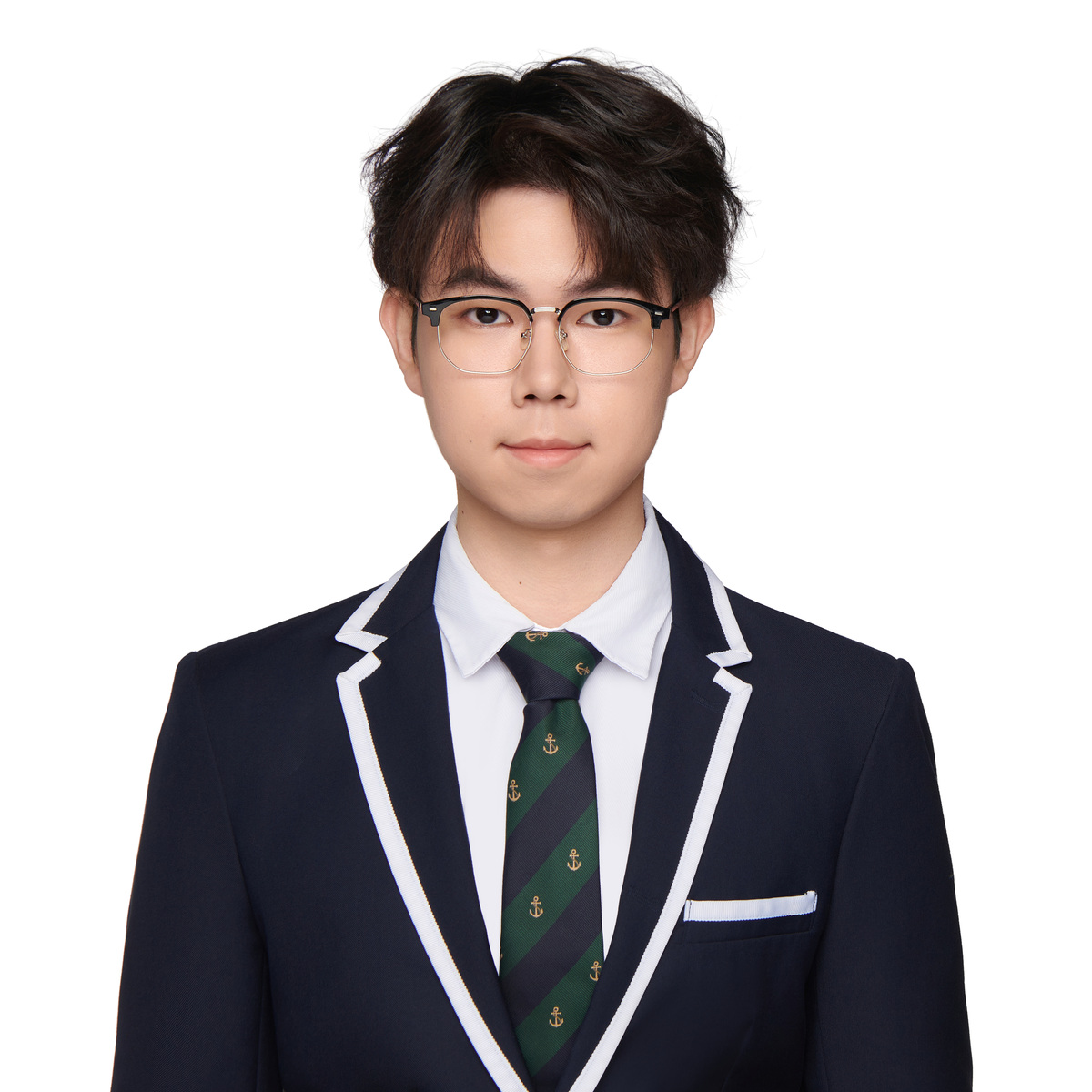}}]{Feifan Xu}
received the bachelor's degree in microelectronics science and engineering
from the School of Integrated Circuits, Sun Yat-sen University. He is currently
a second-year master's student in integrated circuit engineering at the same
school, under the supervision of Prof. Zhang Yueli. His current research focuses
on AI accelerator design, particularly the FPGA-based acceleration and
deployment of large language models.
\end{IEEEbiography}

\begin{IEEEbiography}[{\includegraphics[width=1in,height=1.25in,clip,keepaspectratio]{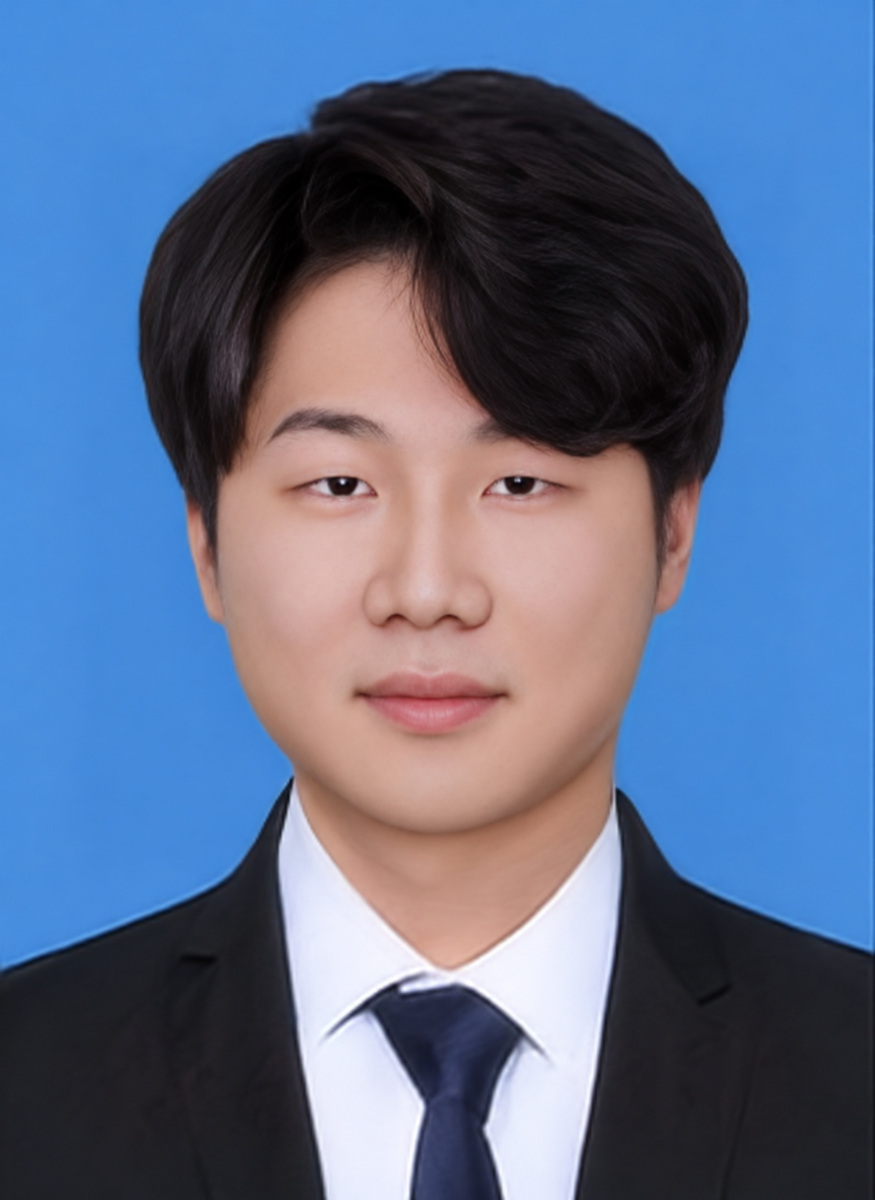}}]{Xingyuan Li}
received the B.S. degree in microelectronics science and engineering from
Sun Yat-sen University, Guangzhou, China, in 2026. He is currently pursuing
the master's degree with the School of Integrated Circuits, Sun Yat-sen
University, Shenzhen, China. His current research interests include
algorithm--hardware co-design for deep learning, large speech models, and
text-to-speech (TTS) synthesis.
\end{IEEEbiography}

\begin{IEEEbiography}[{\includegraphics[width=1in,height=1.25in,clip,keepaspectratio]{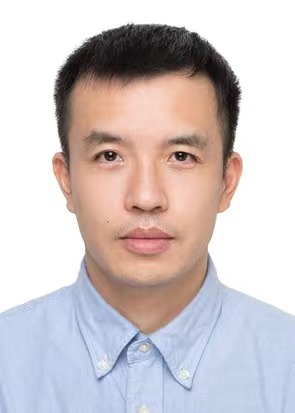}}]{Xinpeng Xing}
(Member, IEEE) received the bachelor's degree in electronics engineering from
Beihang University (BUAA), Beijing, China, in July 2005, the master's degree
in electronics engineering from Tsinghua University, Beijing, China, in July
2008, and the Ph.D. degree in engineering from KU Leuven, Leuven, Belgium,
in October 2013. He was previously with the Shenzhen International Graduate
School (SIGS), Tsinghua University, Shenzhen, China. Since December 2022, he
has been an associate professor with the School of Integrated Circuits,
Sun Yat-sen University, Guangzhou, China. His research interests include
analog and mixed-signal integrated circuit design.
\end{IEEEbiography}

\begin{IEEEbiography}[{\includegraphics[width=1in,height=1.25in,clip,keepaspectratio]{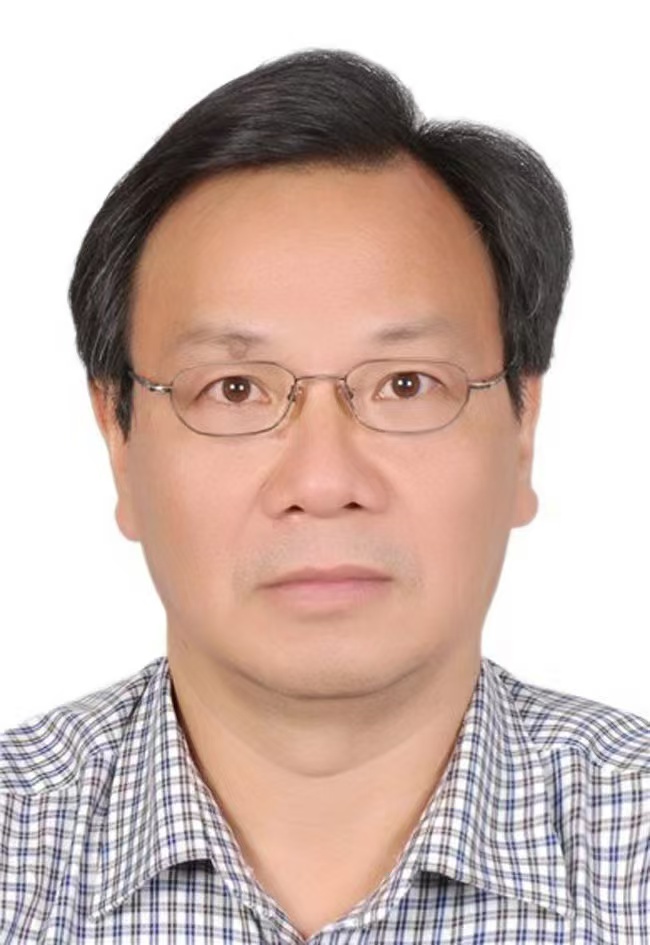}}]{Dihu Chen}
received the B.Sc. and M.Phil. degrees in semiconductor physics from Sichuan
University in 1986 and 1989, respectively, and the Ph.D. degree in solid-state
electronics from the Department of Electronic Engineering, The Chinese
University of Hong Kong, in December 2000. He joined the Department of Physics
at Sun Yat-sen University in July 1986 and became an associate professor in 1995.
From 2000 to 2001, he was a postdoctoral fellow at The Chinese University of
Hong Kong. He subsequently returned to the Department of Microelectronics at
Sun Yat-sen University. He is currently a professor with the School of Integrated
Circuits, Sun Yat-sen University, and serves as director of the Guangdong
Integrated Circuit Technology Engineering Research Center. His current research
interests include electronic devices, IC design methodology, AI algorithms,
and accelerators.
\end{IEEEbiography}

\vfill
\endgroup
\end{document}